\documentclass[aps,twocolumn]{revtex4}

\usepackage{graphicx,amssymb,amsmath,color}

\newcommand{\be}{\begin{equation}}
\newcommand{\ee}{\end{equation}}

\newcommand{\bea}{\begin{eqnarray}}
\newcommand{\eea}{\end{eqnarray}}

\newcommand{\br}{\mathbf{r}}
\newcommand{\bR}{\mathbf{R}}
\newcommand{\bu}{\mathbf{u}}
\newcommand{\bx}{\mathbf{x}}
\newcommand{\bX}{\mathbf{X}}

\newcommand{\bn}{\mathbf{n}}

\newcommand{\bq}{\mathbf{q}}

\newcommand{\md}{\mathrm{d}}

\newcommand{\bt}{\mathbf{t}}
\newcommand{\tnabla}{\tilde{\nabla}}

\def\eq#1{eq.~(\ref{#1})}
\def\eqs#1#2{eqs.~(\ref{#1},\ref{#2})}

\def\fig#1{fig.~\ref{#1}}

\usepackage{array}
\newcolumntype{C}[1]{>{\centering\let\newline\\\arraybackslash\hspace{0pt}}m{#1}}
\newcolumntype{L}[1]{>{\raggedright\let\newline\\\arraybackslash\hspace{0pt}}m{#1}}
\newcolumntype{R}[1]{>{\raggedleft\let\newline\\\arraybackslash\hspace{0pt}}m{#1}}

\begin{document}

\title{Fluctuation tension and shape transition of vesicles:\\ renormalisation calculations and Monte Carlo simulations}

\author{Guillaume Gueguen}
\author{Nicolas Destainville}
\author{Manoel Manghi}
\email{manghi@irsamc.ups-tlse.fr}
\affiliation{Laboratoire de Physique Th\'eorique, Universit\'e de Toulouse, CNRS, UPS, France}

\date{\today}

\begin{abstract}
It has been known for long that the fluctuation surface tension of membranes $r$, computed from the height fluctuation spectrum, is not equal to the bare surface tension $\sigma$, which is introduced in the theory either as a Lagrange multiplier to conserve the total membrane area or as an external constraint. In this work we relate these two surface tensions both analytically and numerically. They are also compared to the Laplace tension $\gamma$, and the mechanical frame tension $\tau$.
Using the Helfrich model and one-loop renormalisation calculations, we obtain, in addition to the effective bending modulus $\kappa_{\rm eff}$, a new expression for the effective surface tension $\sigma_{\rm eff}=\sigma - \epsilon k_{\rm B}T/(2a_p)$ where $k_{\rm B}T$ is the thermal energy, $a_p$ the projected cut-off area, and $\epsilon=3$ or 1 according to the allowed configurations that keep either the projected area or the total area constant. Moreover we show that the crumpling transition for an infinite planar membrane occurs for $\sigma_{\rm eff}=0$, and also that it coincides with vanishing Laplace and frame tensions.
Using extensive Monte Carlo (MC) simulations, triangulated membranes of vesicles made of $N=100-2500$ vertices are simulated within the Helfrich theory. As compared to alternative numerical models, no local constraint is applied and the shape is only controlled by the constant volume, the spontaneous curvature and $\sigma$. It is shown that the numerical fluctuation surface tension $r$ is equal to $\sigma_{\rm eff}$ both with radial MC moves ($\epsilon=3$) and with corrected MC moves locally normal to the fluctuating membrane ($\epsilon=1$). 
For finite vesicles of typical size $R$, two different regimes are defined: a tension regime for $\hat \sigma_{\rm eff}=\sigma_{\rm eff}R^2/\kappa_{\rm eff}>0$ and a bending one for $-1<\hat \sigma_{\rm eff}<0$. A shape transition from a quasi-spherical shape imposed by the large surface energy, to more deformed shapes only controlled by the bending energy, is observed numerically at $\hat \sigma_{\rm eff}\simeq 0$. We propose that the buckling transition, observed for planar supported membranes in the literature, occurs for $\hat \sigma_{\rm eff}\simeq-1$, the associated negative frame tension playing the role of a compressive force. Hence, a precise control of the value of $\sigma_{\rm eff}$ in simulations cannot but enhance our understanding of shape transitions of vesicles and cells.
\end{abstract}

\maketitle

\section{Introduction}

A vesicle, made of a closed lipid bilayer in water, is one of the simplest objects formed by self-assembly of lipids in water (see for instance the book~\cite{giant_vesicles}). Contrary to soap bubbles made of water films stabilised in the air by the two interfaces saturated in surfactants, their bending modulus is much larger, on the order of 10 to $50~k_{\rm B}T$, where $k_{\rm B}T$ is the thermal energy (at room temperature). Hence the physics of vesicles is more intricate and once the microscopic degrees of freedom of the lipids that constitute the membrane are integrated out, the main physical forces that govern the vesicle thermodynamics are the bending forces of the bilayer, its surface tension, and the pressure difference across the membrane.

Despite its apparent simplicity, defining the surface tension of a vesicle is not an easy task and has been highly debated in the last decade. 
Depending on the type of experiment, three surface tensions can be measured (see for instance the review by Bassereau \textit{et al.}~\cite{bassereau} and Fig.~3 therein). The first one is the Laplace tension, $\gamma$, which enters in the Laplace equation for quasi-spherical vesicles. It is similar to the surface tension of bubbles and comes from a pressure difference, as put forward in micropipette aspiration experiments~\cite{evans}. The Laplace surface tension is imposed on the vesicle by sucking a part of its area into the micropipette and therefore stressing it.

The second one is the fluctuation tension $r$ which is extracted from the membrane's fluctuation spectrum at low wavevectors following $\langle |\hat h(q)|^2 \rangle=k_{\rm B}T/[rq^2+\mathcal{O}(q^4)]$. It has been measured e.g. in flickering spectroscopy experiments~\cite{pecreaux,bassereau} and in recent experiments by reflection interference contrast microscopy on adherent membranes~\cite{schmidt} or by dynamic optical displacement spectroscopy~\cite{monzel}.

The third one is the mechanical frame tension $\tau$ which is the tension exerted on planar membranes supported on a frame~\cite{david91,faragoPincus1}. These three surface tensions are summarised in Table~\ref{table1}.
\begin{table*}
\begin{center}
\begin{tabular}{|c|c|c|c|}
\hline
Symbol & Name & Definition & Ref. or eq.\\
\hline
$\gamma$ & Laplace surface tension & enters in the Laplace law  & \eq{Laplace} \\
$r$ & fluctuation surface tension & fitted from the fluctuation spectrum & Ref.~\cite{bassereau}\\
$\tau$ & frame mechanical surface tension & applied by the operator on the supporting frame & Ref.~\cite{david91} \\
$\sigma$ & bare surface tension & enters the Helfrich Hamiltonian & \eq{H2} \\
$\sigma_{\rm eff}$ & effective surface tension & theoretical renormalised fluctuation tension  & \eq{sigma_eff_eps} \\
\hline
\end{tabular}
\end{center}
\caption{Definition of the surface tensions studied in this work together with the relevant references or equations.}\label{table1}
\end{table*}

The relation between these three experimental surface tensions is still debated. In this work we show how they are related to the bare surface tension $\sigma$ entering in the Helfrich theory. In particular we show that when the fluctuation tension vanishes, a vesicle shape transition occurs from a quasi-spherical shape to oblate or prolate shapes.\\

Several theoretical works have already tackled this issue.
By using thermodynamical arguments, Diamant showed in 2011~\cite{diamant11} that two tension variables arise, the Langmuir surface pressure $\Pi$ of the bilayer which is the 2D analogue of the internal pressure of a 3D system, and the Laplace tension $\gamma$ which enters in the Laplace equation and therefore balances the normal stress. Interestingly, Diamant argued that it is not necessary to impose any constraint such as a fixed vesicle area $A$ or a fixed surface tension $\sigma$, and that area relaxation leads to $\Pi+\gamma=0$. The area relaxes with the given constraints of the vesicle volume $V$ (or the pressure difference), the number of molecules inside the vesicle and the number of lipids (or their chemical potential) composing the bilayer. 
Furthermore, Diamant showed that the frame tension $\tau$ is the analogue for an open supported membrane of $\gamma$ for a closed membrane~\cite{diamant11}. It comes from the constraint that the membrane is supported on a projected area $A_{\rm p}$ and they are simply related according to $A_{\rm p}\tau=A\gamma$ (for an incompressible membrane) as discussed below.
Clearly, this work shed light on the definition of the vesicle surface tension. 

However, its drawback comes from the use of classical thermodynamics. No quantitative expressions of $\gamma$ or $\Pi$ can be obtained without the description of the microscopic details of the lipids and their interactions. Doing this requires the use of statistical mechanics but dealing with both the external reservoir which fixes the external pressure, the membrane and the inside of the vesicle is out of reach without introducing a certain level of coarse-graining. Likewise simulating numerically such a system (using for instance molecular dynamics) can only be done on a few microseconds, preventing us to study the various equilibrium shapes.

Physicists therefore model a vesicle as a closed continuous and infinitely thin membrane, embedded in a continuous solvent, a model which does not fulfil the constraints leading to Diamant's result. To fix the area of the membrane, a constraint is therefore introduced: either $A$ is fixed~\cite{seifert1995,milner} but this leads to complicated calculations, or a Lagrange multiplier, the surface tension $\sigma$, is introduced. These types of models have been successfully applied to obtain the mean shape of vesicles by adding a constraint on the global mean curvature~\cite{seifert1997}. The thermal fluctuations around their mean shape have also been studied~\cite{peterson,zhong-can89}. Seifert showed that these two constraints, fixed area or fixed surface tension, led to the same results~\cite{seifert1995}.

Whenever a fixed surface tension $\sigma$ is introduced, the question of its microscopic origin arises. On one hand, the area can change due to the variation of the area per lipid, $a=A/N_{\rm lip}$, at constant number of lipids $N_{\rm lip}$, with a optimum area $a^*$ such that $\Pi(a^*)=0$, following the ideas of Schulman~\cite{schulman,PGG82}. At quadratic order, a phenomenological surface free energy is introduced $f=K(a-a^*)^2/(2a^*)$ thus yielding a simple expression for the surface pressure $\Pi=K (1-a/a^*)$. The values of the compression modulus $K$ and $a^*$ depend on the interactions between lipids. With this approach, the vesicle is viewed as an elastic medium. On the other hand, the area variation can be associated with the addition or removal of lipids from the bilayer at constant $a$ (incompressible membrane) which depends on the chemical potential $\mu$ of adding a lipid from the solution to the bilayer. This is the classical picture for the surface tension between fluids~\cite{Rowlinson_Widom}. To analyse micropipette experiments, both contributions have been simply added in the fitting formula~\cite{faragoPincus1}. 

To obtain analytical expressions for these surface tensions, the Canham-Helfrich theory~\cite{canham,helfrich1973} is used in which the bending energy is quadratic in the average curvature. Hence integrating out the bending fluctuations leads to an effective surface tension $\sigma_{\rm eff}$ which is lowered, compared to the bare surface tension $\sigma$ introduced to control the total area of the membrane. 

From renormalisation arguments, one can identify $\sigma_{\rm eff}$ with the fluctuation tension $r$.
The renormalisation of the Helfrich Hamiltonian for planar membranes has been extensively studied in the last thirty years. Note that Helfrich~\cite{helfrich86} and Kleinert~\cite{kleinert86} used the same renormalisation techniques for vesicles. Several contradictory results have been obtained. Peliti and Leibler~\cite{peliti85} did the first renormalisation calculation for both $\kappa$ and $\sigma$ but focused essentially on $\kappa$. They showed that the bending modulus is lowered by a logarithmic term due to thermal fluctuations. Hence if the effective surface tension vanishes, a crumpling transition occurs for very large membranes, larger than the de Gennes-Taupin persistence length, which increases exponentially with $\kappa/k_{\rm B}T$~\cite{PGG82}. This is very similar to semi-flexible polymers which adopt a random coil when their size is much larger than their persistence length. Although there is a long debate in the literature concerning the numerical prefactor of the corrective term in the renormalized $\kappa$~\cite{helfrich87,forster87,david88}, the formula found by Helfrich and Peliti and Leibler is now generally accepted and has been confirmed numerically~\cite{gompper96}.

The renormalisation of the surface tension has also been discussed by several authors~\cite{peliti85,meunier87,david91,membrane_book_David} but a consensual result is still lacking. Interestingly the renormalisation seems to introduce a corrective term linear in the number of molecules, which yields a much more important correction than for the bending modulus. Cai \textit{et al.}~\cite{cai94} made a rigorous derivation, very similar to the one previously done by Meunier~\cite{meunier87}, and discussed the introduction of a Faddeev-Popov corrective term plus a weak non-linear corrective term. The Faddeev-Popov correction is non-negligible according to Cai \textit{et al.} whereas David~\cite{membrane_book_David} argued that it does not introduce any corrective term in the effective surface tension. No decisive answer is obtained concerning the renormalisation flow equations and the validity of the computed effective surface tension.
David and Leibler~\cite{david91} discussed the different regimes depending on the dominating mechanism, either the tension, the rigidity or the thermal fluctuations, but obtained expressions for $\sigma_{\rm eff}$ different from the ones by Cai \textit{et al.} and Meunier. The method of renormalisation calculation also differs which makes the comparison difficult. Meunier used the classical Wilson procedure~\cite{meunier87}, other authors use the effective potential~\cite{peliti85,cai94}, and others the ``background method''~\cite{forster86,kleinert86,polyakov86,david91}. A precise calculation of the renormalisation flow equations and their exact solution in the different regimes is still missing. 

The frame tension, defined as the free energy that must be supplied to increase the \textit{projected} area of the membrane on a planar frame by one unit, has only been studied in the Gaussian approximation. Even though it was initially computed for planar membranes using the stress tensor~\cite{fournierPRL}, Barbetta and Fournier showed that it can be extended to quasi-spherical vesicles~\cite{barbetta}, the projected area then being the area of the sphere of same volume.

How are the fluctuation tension and the frame tension related? Using 1D or 2D simulations of planar membranes,  some recent works argue that they are equal~\cite{farago,schmid,avital,shiba}. Other works found using coarse-grained simulations~\cite{imparato} and analytical arguments~\cite{fournierPRL} that they are different (see also the recent review~\cite{deserno}). These studies essentially focus on the value of the residual bare surface tension for which the fluctuation tension vanishes using the Gaussian approximation. But at such low surface tensions, the renormalisation of the fluctuation and the frame tensions must be properly taken into account.
Moreover no quantitative comparison between the effective surface tension obtained by renormalisation calculations and the fluctuation tension measured in numerical simulations has been done in the literature.

\begin{figure}
\centering
\includegraphics[width=0.99\linewidth]{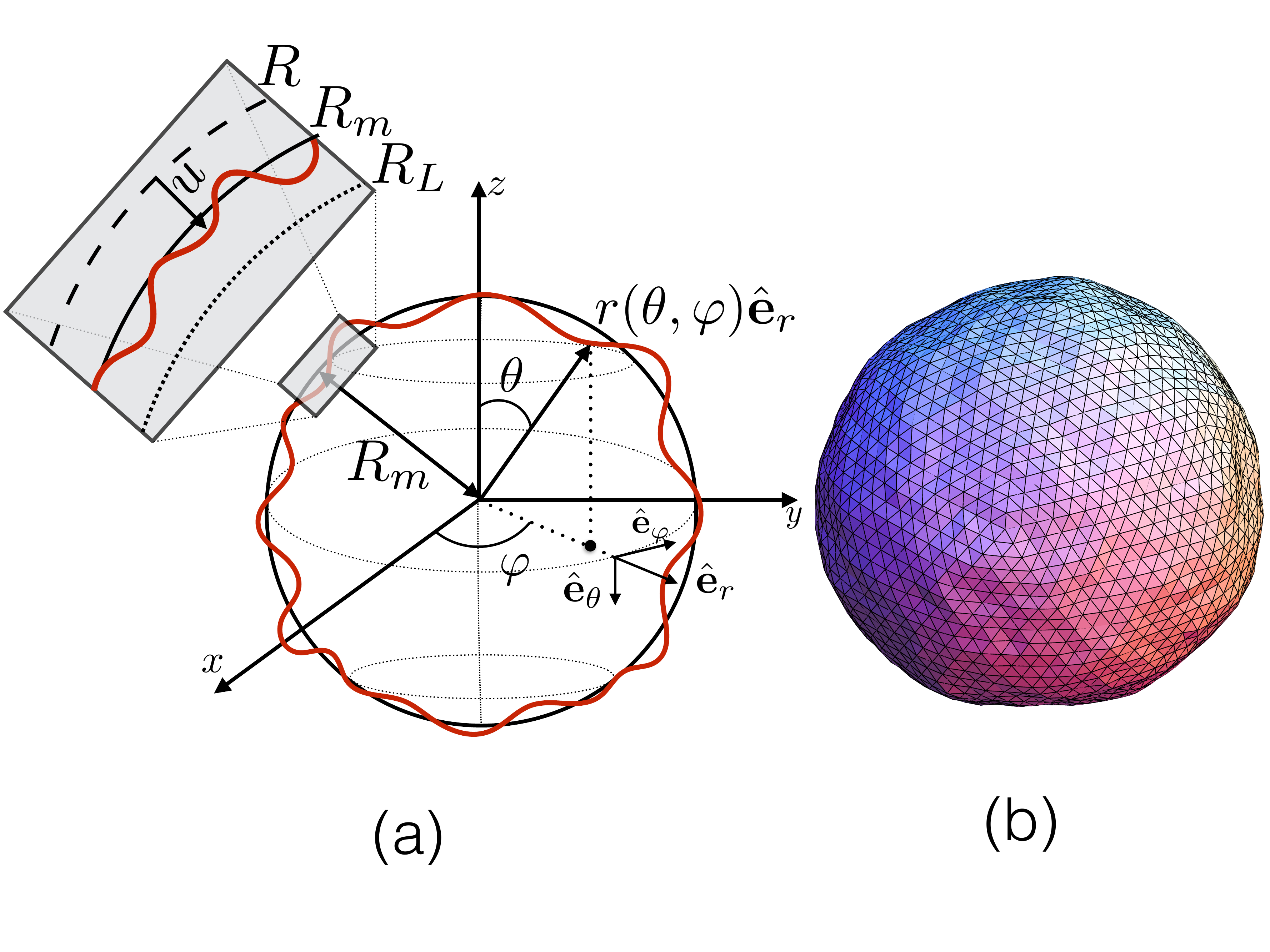} 
\caption{(a)~Sketch of a fluctuating quasi-spherical vesicle of mean radius $R_m$. The position of the membrane is given by $\br=R[1+u(\theta,\phi)]\hat{\mathbf{e}}_r$ where $R=[3V/(4\pi)]^{1/3}>R_m$ and $V$ is the vesicle volume. The ``volume'' radius $R$ and the Laplace radius $R_L$ are shown in the zoom. (b)~Snapshot of a simulated vesicle with  $N=2562$ vertices ($\beta\kappa=10$, $C=0$, $R^2\sigma/\kappa=60$, see text).}\label{fig_intro}
\end{figure}

The aim of this work is threefold. First, by performing renormalisation calculations, we compute the three renormalized fluctuation, Laplace and frame surface tensions. In particular we show that the fluctuation and the Laplace tensions are equal (when the Faddeev-Popov corrective term is taken into account). Second we perform Monte Carlo simulations of a vesicle with global constraints (fixed volume, fixed area or fixed surface tension, see~\fig{fig_intro}) and without any local constraints between the vertices of the triangulated surface, as it is done in other numerical works~\cite{sunilkumar1997,hu2011,amazon2013,amazon2014}. By extracting the fluctuation surface tension from the numerical height-height correlation functions, we confirm our analytical results. Finally we identify the shape transition of the vesicle when the effective surface tension is significantly decreased.

The paper is organised as follows. In Section~\ref{Gauss}, we review the calculation of the Laplace and frame surface tensions in the framework of the Helfrich theory for quasi-spherical vesicles. In particular, we show that in the limit of very large vesicle radii, we recover the results for planar membranes in the Gaussian approximation. 
The renormalisation equations for the bending modulus and the surface tension are derived in Section~\ref{renormalisation} for a planar membrane. The effective surface tension is deduced with and without the corrective Faddeev-Popov term and the renormalized Laplace end frame tensions are then computed. Section~\ref{Num_Meth} is devoted to the numerical methods of the Monte Carlo vesicle simulations. Section~\ref{Num_Res} contains the numerical results on the fluctuation tension which compares satisfactorily to our analytical formula. The shape transition of the simulated vesicles is then studied as a function of the bending modulus and the spontaneous curvature. We close by discussing several results on planar membrane in the literature which confirm our findings.

\section{Analytical considerations on quasi-spherical vesicles}
\label{Gauss}

\subsection{Laplace surface tension in two thermodynamic ensembles}

According to Helfrich~\cite{helfrich1973} and Seifert~\cite{seifert1997} the volume $V$ and the area $A$ of a vesicle can be considered as constant. They related it to the fact that the membrane is viewed as impermeable and incompressible and there are almost no free lipids in the surrounding medium which could form a reservoir of membrane particles. As shown by Diamant using classical thermodynamics~\cite{diamant11}, one derives the Laplace formula for any closed vesicle:
\be
\Delta p \equiv p_{\rm in}-p_{\rm out}= \frac2{R_{\rm L}}\gamma \quad \mathrm{where}\quad R_{\rm L}\equiv \frac{3V}A
\label{Laplace}
\ee
where $p_{\rm in}$ (respectively $p_{\rm out}$) is the pressure inside (resp. outside) the vesicle, and $\gamma$ is called the \textit{Laplace surface tension}. 
This tension is the actual surface tension of a vesicle which can be measured using, for instance, micropipette experiments~\cite{evans}. 
The Laplace radius $R_{\rm L}$ can be measured and, in general does not coincide exactly with the mean vesicle radius $R_m$ or the radius $R$ of the sphere having the same volume (see \fig{fig_intro}). They do coincide when the area $A$ is equal to the area of the sphere having the same volume $A_{\rm s}\equiv3V/R=4\pi R^2$ i.e. only when the vesicle is exactly a sphere. In general, due to thermal fluctuations, $A> A_{\rm s}$ and $R_{\rm L}<R$. One defines the relative excess area as
\be
\alpha \equiv \frac{A-A_{\rm s}}{A_{\rm s}}
\label{alpha}
\ee
Bending modes are excited by thermal fluctuations which renders the spherical vesicle more fuzzy and the vesicle radius not easy to measure.

The total free energy of a system of volume $V_0$ composed by a vesicle, made of $N_{\rm lip}$ lipids, floating in a fluid, is
\be
F = F_{\rm in}(V)+F_{\rm out}(V_0-V)+F_{\rm memb}(A,V,N_{\rm lip})
\label{Ftot}
\ee
where the free energy of the fluid inside (respectively outside) the vesicle is noted $F_{\rm in}$ (resp. $F_{\rm out}$).
The membrane free energy, in this $(A,V,N_{\rm lip})$ ensemble, is $F_{\rm memb}$.

Since $V$ and $A$ are internal constraints, thermodynamic equilibrium is obtained when $F$ is minimal with respect to variations in $V$ and $A$. Hence $\left(\partial F/\partial V\right)_{A,N_{\rm lip}}=0$ yields \eq{Laplace} where $p_{\rm out} =-\partial F_{\rm out}/\partial V$ (the same equality holds for $p_{\rm in}$) and the Laplace tension is given by
\be
\gamma =  \frac32 v \left(\frac{\partial f}{\partial v}\right)_a
\ee
Here we have written $F_{\rm memb}=A f(a,v)$ where 
\be
a\equiv \frac{A}{N_{\rm lip}};\qquad v\equiv 6\sqrt{\pi}\frac{V}{A^{3/2}}
\label{av}
\ee
are respectively the area per lipid and the reduced volume.
$\left(\partial F/\partial A\right)_{V,N_{\rm lip}}=0$ yields similarly
\be
\gamma=f + a \left(\frac{\partial f}{\partial a}\right)_v
\ee
Multiplicating by $A$, we recover the classical formula~\cite{Rowlinson_Widom} $F_{\rm memb}=\gamma A + \mu N_{\rm lip}$ where $\mu =\left(\partial F_{\rm memb}/\partial N_{\rm lip}\right)_{A,V}=-a^2\left(\partial f/\partial a\right)_v$ is the chemical potential of the lipids composing the vesicle membrane
\footnote{Note that in the classical Laplace derivation for geometric spheres, $F_{\rm memb}= \gamma A$ is independent of $V$, but to keep $A$ and $V$ as independent variables, a Lagrange multiplier $\lambda$ is introduced to enforce the spherical constraint  $v=1$ by adding for instance $\lambda(6\sqrt{\pi}V-A^{3/2})$ to the free energy. By minimising and eliminating $\lambda$, we still recover \eq{Laplace}.}.

The membrane free energy is given by $F_{\rm memb}=- k_{\rm B}T \ln \mathcal{Z}$ where the partition function $\mathcal{Z}$ writes
\be
\mathcal{Z} = \int \mathcal{D}\br \,\delta(\mathcal{A}[\br]-A)) \delta(\mathcal{V}[\br]-V)) \exp(-\beta {\cal H}_{\rm h}[\br])
\label{Z}
\ee
The position of the membrane at the point located by the (curvilinear) coordinates $(u^i;i=1,2)$ is $\br(\{u^i\})$. The classical Canham-Helfrich effective Hamiltonian is~\cite{canham,helfrich1973}
\be
{\cal H}_{\rm h}[\br] = \frac\kappa2\int \md u^1\md u^2 \sqrt{g}  \ (2H - C)^2  \label{H0}
\ee
where $\kappa$ is the bending modulus and $C$ the spontaneous curvature. The mean curvature $2H=g^{ij}K_{ij}$ is the trace of the curvature tensor $K_{ij}$ which can be defined by $\partial_i\bn=K_{ij}\bt^{j}$ where $\bt_i=\partial_i\br$ and $\bn$ is the normal vector to the surface. The metric is defined as $g_{ij}=\partial_i\br \,\partial_j\br$ and $g=\det g_{ij}$ is its determinant. The functional
\be
\mathcal{A}[\br]=\int \md u^1\md u^2 \sqrt{g}
\ee
is the vesicle area and $\mathcal{V}[\br]$ its volume, both written as a function of the $\br$. Their exact expression will be given later for a given parametrisation $u^i$ of the membrane.

In most theoretical works, it is extremely difficult to introduce the hard constraint $A=\mathrm{const.}$, which is thus controlled \textit{via} a Lagrange multiplier $\sigma$, homogeneous to a surface tension, but which is not \textit{a priori} an experimentally measurable one. 
Keeping the hard constraint for the volume, the new partition function is
\bea
\mathcal{Q} &=& \int_0^\infty \md A e^{-\beta \sigma A} \mathcal{Z}(A)\label{defQ} \nonumber\\
&=& \int \mathcal{D}\br  \, \delta(\mathcal{V}[\br]-V) \ e^{-\beta ({\cal H}_{\rm h}[\br] + \sigma \mathcal{A}[\br])}
\label{Q}
\eea
The associated membrane free energy, noted $G_{\rm memb}=-k_{\rm B}T \ln\mathcal{Q}$, now depends on $(\sigma,V,N_{\rm lip})$ and the average area noted $A$ is given by
\be
A=\langle \mathcal{A}[\br]\rangle=\left(\frac{\partial G_{\rm memb}}{\partial \sigma}\right)_V
\label{defA0}
\ee
which relates implicitly $\sigma$ to $A$.
In this new statistical ensemble, the total free energy in \eq{Ftot} is modified by replacing $F_{\rm memb}$ by $G_{\rm memb}$. Following the same approach as above, one obtains again the Laplace equation~(\ref{Laplace}), with $R_{\rm L}$ replaced by $\langle R_{\rm L}\rangle=3V/\langle \mathcal{A}[\br]\rangle$:
\be
\gamma = \frac32\frac{V}{A} \left(\frac{\partial G_{\rm memb}}{\partial V} \right)_\sigma
\label{def_gamma}
\ee
Writing $G_{\rm memb}=A_{\rm s}\, g(\sigma, a_{\rm s})$ where $A_{\rm s}=(4\pi)^{1/3}(3V)^{2/3}$ is the area of the hypothetical sphere of volume $V$ and $ a_{\rm s}=A_{\rm s}/N_{\rm lip}$, one has
\be
\gamma= \frac{A_{\rm s}}{A}\left[g+a_{\rm s}\left(\frac{\partial g}{\partial a_{\rm s}}\right)_\sigma \right]
\label{gamma_nocurv}
\ee
In the following we compute the Laplace tension of quasi-spherical vesicles ($\alpha\ll 1$).

\subsection{Expansion around a spherical vesicle}

It is interesting to rewrite the total Hamiltonian ${\cal H}={\cal H}_{\rm h}+\sigma {\cal A}$ appearing in \eq{Q} [where ${\cal H}_{\rm h}$ is defined in \eq{H0}] as
\bea
{\cal H} &=& \frac\kappa2\int \md{\cal A}  \  \left(2H - \frac2R\right)^2 + \bar \sigma {\cal A} \nonumber\\
&& +\kappa \left(\frac2R-C\right) \int \md{\cal A} \left(2H - \frac2R\right) \label{H2}
\eea
where $2/R$, with $R\equiv[3V/(4\pi)]^{1/3}$, would be the natural curvature of the vesicle imposed by the volume, and $\bar \sigma$ is an effective surface tension which depends on the spontaneous curvature $C$:
\be
\bar \sigma=\sigma +\frac{\kappa}{2}\left(\frac{2}{R}-C\right)^2
\ee

For a taut vesicle which keeps a quasi-spherical shape, the natural parametrisation $u^i$ of the membrane is the usual spherical coordinates $(u^1=\theta,u^2=\varphi)$. We expand $\br$ around the sphere following $
\br=r(\theta,\varphi)\hat {\bf e}_r=R[1+ u(\theta,\varphi)]\hat {\bf e}_r$ [see~\fig{fig_intro}(a)]. 
Following Refs.~\cite{milner,helfrich86,zhong-can89,seifert1995,GueguenEPJE}, we thus expand ${\cal H}$ up to second order in $u$ with
\bea
\md\mathcal{A} &\simeq& R^2\left[1+2u+u^2+\frac12(\tnabla u)^2 \right]\md\Omega \label{dA}\\
(2H) \md\mathcal{A} &\simeq& 2R\left[1+u-\frac12\tnabla^2u+\frac12(\tnabla u)^2 \right]\md\Omega  \label{2HdA}\\
(2H)^2\md\mathcal{A} &\simeq& 4 \left[1-\tnabla^2u+ u\tnabla^2u\right.\nonumber\\
&& \left. +\frac14(\tnabla^2u)^2+\frac12(\tnabla u)^2 \right]\md\Omega 
\eea
where $\md\Omega= \sin\theta \md\theta\md\varphi$ and $\tilde{\nabla} \equiv \hat {\bf e}_\theta\partial_\theta +\hat {\bf e}_\varphi\frac1{\sin\theta} \partial_\varphi$. We use the decomposition of $u$ in the spherical harmonic basis
\be
u(\theta,\varphi)=\frac{u_{00}}{\sqrt{4\pi}}+\sum_{l=1}^{l_{\max}}\sum_{m=-l}^l u_{lm}Y_l^m(\theta,\varphi)
\label{harmonics}
\ee
where the cutoff $l_{\max}$ is related to the number of modes, $N$, by 
\be
N=\sum_{l=0}^{l_{\max}}(2l+1)=(l_{\max}+1)^2
\ee
As already noted by Farago and Pincus~\cite{faragoPincus1}, since we use the Helfrich Hamiltonian where the membrane is viewed as infinitely thin, the conformations of the surface cannot be defined below some microscopic length scale on the order of the actual membrane thickness ($\simeq5$~nm). The number of modes is therefore fixed according to $N/A_{\rm s}=a_{\rm s}^{-1}$, which is fixed, and each patch of area $a_{\rm s}$ contains around 10 to 100 lipids, a number which depends on the total membrane area $A$. Hence the present theory does not consider any lipid degrees of freedom (such as translation or rotation).
In this basis, the Hamiltonian is rewritten as
\be
{\cal H}=E_{\rm sph}(R)+ \frac{\kappa}2 \sum_{l=1}^{l_{\max}}\sum_{m=-l}^l  H(l) |u_{lm}|^2 
\label{H2_sigV}
\ee
where
\bea
E_{\rm sph}(R) &=& 4\pi R^2 \bar \sigma\\
H(l) &=& [l(l+1)-2]\left[l(l+1)-2+ \hat \sigma \right]\label{Hl}
\eea
where
\be
\hat\sigma \equiv \frac{R^2\bar\sigma}\kappa=\frac{R^2\sigma}\kappa+\frac12(2-RC)^2
\label{sigmahat}
\ee 
The result \eq{H2_sigV} has been obtained by many authors~\cite{milner,helfrich86,zhong-can89,barbetta,GueguenEPJE}.
Note that to obtain \eq{Hl}, the following condition of constant volume has been used
\bea
V=\frac{4\pi R^3}3&=&\frac{R^3}3\int\md \Omega(1+u)^3\\
&=& \frac{R^3}3\int\md \Omega[1+3u+3u^2+\mathcal{O}(u^3)]\label{expa}
\eea
which implies that 
\be
u_{00}=-\frac1{\sqrt{4\pi}}\sum_{l=1}^{l_{\max}}\sum_{m=-l}^l |u_{lm}|^2
\label{Vconserv}
\ee 
at this order. This is the mathematical origin of the factor $l(l+1)-2$ in \eq{Hl}. The sum is taken from $l=1$ since the mode $l=0$, corresponding to the vesicle uniform dilatation/compression, is not counted due to the constant volume condition. Interestingly the 3 soft modes $l=1$, which correspond to the vesicle translation in the 3 directions, do not contribute to the surface and elastic free energy of the vesicle (they lead to a term in $-k_{\rm B}T\ln V_0$ which we do not consider in the following).

In particular, injecting the volume conservation \eq{Vconserv} in \eqs{dA}{2HdA}, the last term of \eq{H2} vanishes at order 2 in $u$. Hence, the effect of the spontaneous curvature $C$ is solely to increase the bare surface tension $\sigma$. Indeed, if the prescribed radius $R$ does not match the spontaneous radius of curvature $2/C$, a supplemental bending energy per unit surface arises.

The free energy per unit surface $A_{\rm s}$ is
\be
g=\bar \sigma + \frac{k_{\rm B}T}{8\pi R^2} \sum_{l=2}^{l_{\max}}(2l+1)\ln\left[\frac{\beta\kappa d^2}{R^2}H(l)\right] \label{FS}
\ee
where $d$ is a small length which must be introduced to render $\mathcal{Q}$ dimensionless and is on the order of the membrane thickness. It does not play any role in the following calculation of measurable thermodynamic quantities.
Since $l_{\max}\simeq\sqrt{N}$, the 2nd term is proportional to $N/(4\pi R^2)=a_{\rm s}^{-1}$, $g$ in \eq{FS} is effectively a function of $\sigma$ and $a_{\rm s}$. It also depends on three other dimensionless parameters, $\sigma R^2/\kappa$, $\beta\kappa$ and $RC$. 

\subsection{Correlation function and excess area}

The correlation function $\langle u(\br_1)u(\br_2)\rangle$ depends only on the angle $\Theta$ defined by $\br_1\cdot\br_2=\cos\Theta$. One gets
\bea
\mathcal{C}(\Theta;\hat\sigma) &=& \langle u(\Theta)u(0)\rangle \nonumber\\
&=& \frac{k_{\rm B}T}{4\pi\kappa} \sum_{l=2}^{l_{\max}} \frac{(2l+1)P_l(\cos\Theta)}{H(l)} \label{CS}
\eea
where the $P_l(x)$ are the Legendre polynomials.
In particular, the membrane roughness is
\be
\langle u^2\rangle =\mathcal{C}(0;\hat\sigma) = \frac{k_{\rm B}T}{4\pi\kappa} \sum_{l=2}^{l_{\max}} \frac{(2l+1)}{H(l)}
\label{roughS}
\ee
The average area of the membrane is given by \eq{defA0}, which allows us to write the dimensionless excess area (defined in \eq{alpha}) as
\be 
\alpha =\frac{k_{\rm B}T}{8\pi\kappa}\sum_{l=2}^{l_{\max}} \frac{2l+1}{l(l+1)-2 +\hat \sigma}\label{alphaS}
\ee
which is a function of $\hat \sigma$, $\beta\kappa$ and $N$ like the correlation function defined in \eq{CS}.
 
\begin{figure}
\centering
\includegraphics[width=0.99\linewidth]{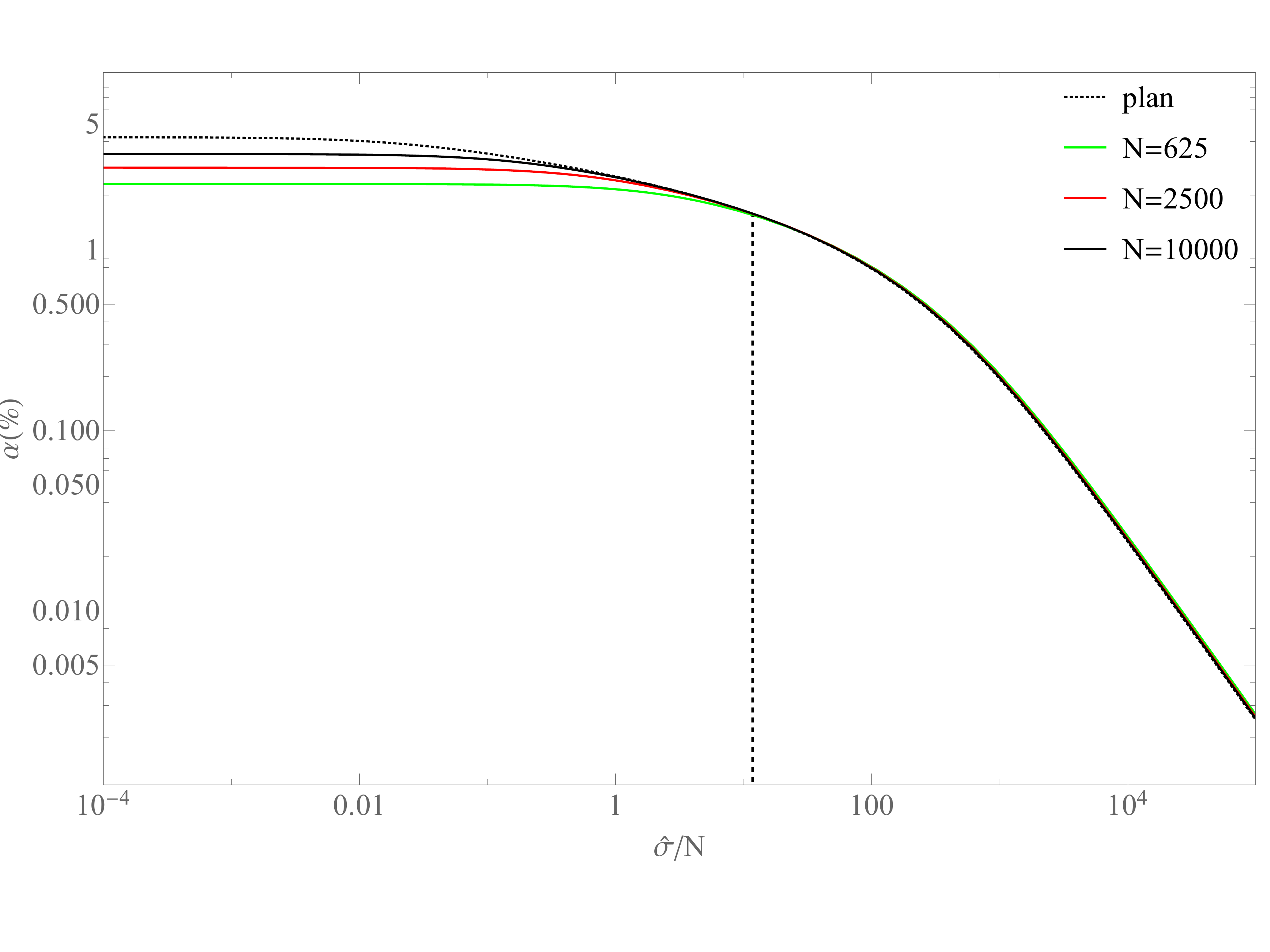} 
\caption{Excess area $\alpha$ vs. the dimensionless surface tension $\hat\sigma/N$ for various values of $N$ ($R=1$, $\beta\kappa=10$, $C=2$) (log-log scale). The dashed vertical line divides the plot in two different regimes: on the right $\hat\sigma$ scales as $N$ whereas on the left $\alpha\simeq \mathrm{const.}$ For the planar case, $\Lambda=200$ (see text). \label{alphag}}
\end{figure}
In Fig.~\ref{alphag} the excess area $\alpha$ [given by \eq{alphaS}] is plotted against $\hat \sigma/N$ for various values of the number of modes $N$ keeping the vesicle area constant ($RC=2$ and $\beta\kappa=10$). We observe that the curves superimpose for large values, $\hat \sigma/N>10$, forming a master curve. Hence, for a fixed excess area $\alpha$, the surface tension scales with the number of degrees of freedom $N$ for large $\sigma$. The excess area is therefore ``absorbed'' by the $N$ modes. When $\sigma$ decreases (or $N$ increases) more and more excess area is created. Finite size effects start to be felt for $\hat \sigma/N<10$. For these low $\sigma$ values, both the roughness and the excess area $\alpha$ saturate and are no more controlled by $\sigma$. We will show below that the specific surface tension value of $\hat \sigma\simeq N$ (see dashed vertical line in the figure) also corresponds to a crossover from a quasi-spherical shape to a shape which is no more spherical, the first instable modes being for $l=2$~\cite{seifert1995,zhong-can89}.

\subsection{Laplace and frame tensions of quasi-spherical vesicles}

The Laplace surface tension is found using \eq{def_gamma}:
\be
\gamma = \bar \sigma -\frac{k_{\rm B}TN'}{2 A} -\frac{\kappa}{(1+\alpha)R}\left(\frac2R-C\right)
\label{gamma}
\ee
where $N'=\sum_{l\geq 2} (2l+1)=N-4$, $A= A_{\rm s} (1+\alpha)=4\pi R^2 (1+\alpha)$ and $\alpha$ is given by \eq{alphaS}. Note that the last term vanishes both for $2/R=C$ and $2/R=0$ (planar case).

The second term in \eq{gamma} is related to the averaged bending energy defined in \eq{H0} through
\bea
\frac{\Delta E_{\rm b}}{A} &=& \frac{\kappa}{2A}\left\langle\int\md \mathcal{A}\left(2H-\frac2R\right)^2\right\rangle \nonumber \\
&=& \frac{k_{\rm B}TN'}{2 A} -\frac{\alpha}{1+\alpha}\bar\sigma \label{Hcurv}
\eea
the last equation coming from the expansion at Gaussian order.

Now we examine the frame tension $\tau$ first introduced in the context of planar membranes standing on a frame~\cite{david91}. It is the elastic tension which, in equilibrium, balances the tension applied by the operator to maintain the projected area constant. The frame tension has been extended to the case of quasi-spherical vesicles~\cite{fournierPRL,barbetta}. However the surface tension depends on the choice of the reference surface~\cite{Rowlinson_Widom}: it can be the sphere having the same surface as the vesicle or the same volume. Following Barbetta \textit{et al.}~\cite{barbetta}, we choose the sphere of volume $4\pi R^3/3$. Its surface is $A_{\rm s}=4\pi R^2$. The frame tension is therefore the averaged stress component along ${\bf e}_\theta$ and its thermodynamical definition is
\be
\tau=  \left(\frac{\partial G_{\rm memb}}{\partial A_{\rm s}} \right)_{N,\sigma}
\label{def_tau}
\ee
Comparing \eqs{def_gamma}{def_tau} one readily finds
\bea
\tau &=& \frac{A}{A_{\rm s}}\gamma = (1+\alpha)\gamma \label{tau00}\\
&=&  (1+\alpha)\bar\sigma - \frac{k_{\rm B}T N'}{2A_{\rm s}} -\frac{\kappa}R \left(\frac2R-C\right)
\label{tau0}
\eea
Eq.~(\ref{tau00}) has first been demonstrated by Diamant~\cite{diamant11} and \eq{tau0} generalises to any value of $C$ the result found by Barbetta \textit{et al.} for $C=0$. 

\begin{figure}
\centering
\includegraphics[width=0.99\linewidth]{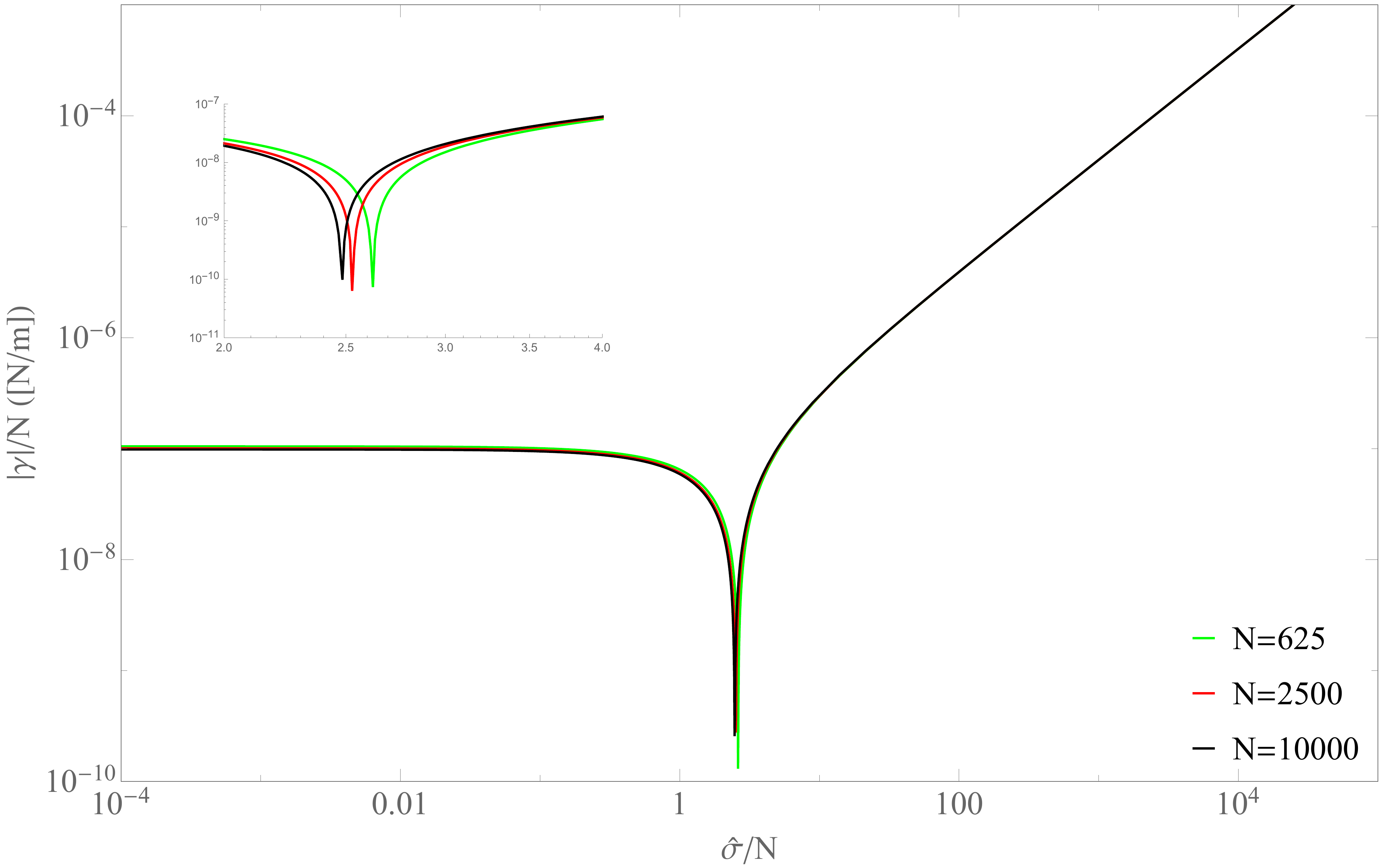} 
\caption{Laplace surface tension per mode $\gamma/N$, given in \eq{gamma}, vs. the dimensionless surface tension $\hat\sigma/N$ for various values of $N$ ($R=1$, $\beta\kappa=10$, $C=2$) (log-log scale). One observes a superimposed linear dependence for $\hat\sigma/N>4$ (see the zoom in the inset) and that $\gamma\to0$ for $\hat\sigma/N\simeq2.5$. For lower values of $\hat\sigma$, $\gamma<0$. The planar case (\eq{gammaplane} with $\Lambda=200$) is also plotted and cannot be distinguished from the $N=10000$ curve. \label{gammag}}
\end{figure}

In \fig{gammag}, the Laplace tension is plotted against the dimensionless surface tension $\hat\sigma/N$. Since the excess area $\alpha<5\%$  for this range of $\sigma$ values, the frame tension $\tau$ has the same behaviour. By comparing \fig{alphag} and \fig{gammag}, one observes that $\gamma\to0$ when $\alpha$ starts to saturate, $\alpha$ being independent of $\sigma$ when $\gamma$ and $\tau$ are negative. What is the physical meaning of $\gamma<0$? From \eq{Laplace} it suggests that the spherical shape of the vesicle is no more stable. Indeed the vesicle does not want to minimise its area any more. Of course the hypothesis of a quasi-spherical vesicle is no more valid and one should go beyond the quadratic expansion in $u$. This will be done in Section~\ref{renormalisation}.

The distinction between the two surface tensions, $\gamma$ and $\tau$ is related to the two radii, the Laplace radius $R_{L}$ defined through \eq{def_gamma} and the radius $R$ related to the vesicle volume.  They are simply related through $R_{L}=R/(1+\alpha)$. A third radius is the mean radius $R_{\rm m}\equiv \langle r\rangle\simeq R(1-\langle u^2\rangle)<R$ since due to the volume conservation, $\langle u\rangle\simeq-\langle u^2\rangle$.  We have numerically checked that $R_L\leq R_{\rm m}\leq R$, the equality occurring only for vanishing membrane fluctuations. Hence one has 
\be
p_{\rm in}-p_{\rm out}= \frac2{R_{\rm L}}\gamma= \frac2{R}\tau
\ee
and we recover the result that the definition of the surface tension depends on the choice of the so-called dividing surface, and they coincide in the limit where the membrane ``thickness'', or more precisely here the membrane roughness, tends to zero~\cite{Rowlinson_Widom}.

\subsection{Planar membrane as the thermodynamic limit of a vesicle}

It is interesting to compute theses values in the \textit{thermodynamic limit} corresponding to $N\to\infty$ and $R\to \infty$ while keeping $a_{\rm s}=4\pi R^2/N=4\pi/\Lambda^2$ constant, where $\Lambda$ is the infrared cutoff in the Fourier space.

In this limit, the modes $(l,m)$ are mapped onto wavevectors $\bq$ contained in the plane $z=0$ with a translational invariance. These wavevectors are discretised following $\bq=2\pi(n_x{\bf \hat x}+n_y{\bf \hat y})/L$ where the projected area is $A_{\rm p}=L^2$ (which is equal to the previous $A_{\rm s}=4\pi R^2$). In the continuous limit and using the 2D isotropy (with $|\bq|=q\leq \Lambda$) by equating the number of modes in a shell of radius $q$, one obtains $2R^2 q \md q=(2l+1)\md l$ which leads to $q^2=[l(l+1)-2]/R^2$. Hence \eq{H2_sigV} is transformed into
\be
\mathcal{H}_0= A_{\rm p} \bar \sigma+ \frac{\kappa}2 \int_0^\Lambda \frac{q\md q}{2\pi}q^2(q^2+\xi^{-2})|\hat h(q)|^2 
\label{H0_pl}
\ee
where $\xi$ is the correlation length such that 
\be
\xi^{-2}=\frac{\bar \sigma}{\kappa}=\frac{\sigma}{\kappa}+ \frac{C^2}2
\label{xi}
\ee
and we recover the planar Helfrich Hamiltonian in the Fourier space. In real space it writes
\be
\mathcal{H}_0= A_{\rm p} \bar \sigma+ \frac12 \int_0^\infty \md \bx \left[\bar \sigma\left(\nabla h(\bx)\right)^2 + \kappa\left(\nabla^2 h(\bx)\right)^2 \right] \label{H1}
\ee
Note that in this limit we keep $\bar\sigma=\sigma +\kappa C^2/2$ (since $2/R\to0$) which depends on the spontaneous curvature $C$, but the average curvature $\langle \nabla^2 h\rangle=0$ at this order (the last term of \eq{H2} vanishes) as usual  in any planar Hamiltonian and first noticed by Meunier~\cite{meunier87}.

The free energy per unit surface \eq{FS} becomes
\bea
g_{\rm pl} &=& \bar\sigma + \frac{k_{\rm B}T}{4\pi}\int_0^{\Lambda} q\md q \ln\left[\beta\kappa d^2 R^2q^2(q^2+\xi^{-2}) \right] \nonumber \\
&=& \bar\sigma +  \frac{k_{\rm B}T \Lambda^2}{8\pi}\left\{ \ln\left(\beta\bar\sigma d^2 \Lambda^2 R^2 \right)-2\right.\nonumber \\
&&\left. + \left[1+(\xi \Lambda)^{-2}\right]\ln\left[1+(\xi \Lambda)^2\right] \right\}
\label{fplane}
\eea
which corresponds exactly to the expression given in Ref.~\cite{schmid}.
The excess area is obtained using \eq{defA0}:
\be
\alpha_{\rm pl} = \frac{k_{\rm B}T}{8\pi \kappa} \ln\left[1+(\xi \Lambda)^2\right] 
\label{alphaP}
\ee
Similarly, the roughness is
\be
\langle h^2\rangle_c = \frac{k_{\rm B}T}{2\pi\kappa}\int_{\pi/L}^{\Lambda} \frac{q\md q}{ q^2(q^2+\xi^{-2})}= \frac{k_{\rm B}T}{4\pi\sigma}\ln\left[\frac{L^2\Lambda^2/\pi^2}{1+(\xi\Lambda)^2}\right]
\label{roughP}
\ee

The Laplace tension defined in \eq{gamma} does not change and is now
\be
\gamma = \sigma+\frac12 \kappa C^2 -\frac{k_{\rm B}T\Lambda^2}{8\pi (1+\alpha_{\rm pl})}
\label{gammaplane}
\ee

The frame tension is the conjugated variable to projected area $A_{\rm p}$ for planar membranes (normally by keeping fixed $A$ but this condition is generally relaxed, see \cite{diamant11}). It can be computed in both ways: either by differentiating the total free energy by $A_{\rm p}$ and taking the thermodynamic limit $A_{\rm p}\to\infty$ in a second step, which leads to the same quantity obtained using the stress tensor, or by differentiating the free energy $G_{\rm pl}$ in the thermodynamic limit with respect to $A_{\rm p}$ while keeping the number of modes $N$ constant. The two calculations lead to the same result, contrary to what was suggested in Ref.~\cite{barbetta}. One finds
\bea
\tau &=& \left(\frac{\partial G_{\rm pl}}{\partial A_{\rm p}}\right)_N= -\Lambda^4 \left(\frac{\partial g_{\rm pl}/\Lambda^2}{\partial \Lambda^2}\right)_N  \nonumber\\
&=& (1+\alpha_{\rm pl}) \gamma\\
&=&  (1+\alpha_{\rm pl}) \bar \sigma  - \frac{k_{\rm B}T\Lambda^2}{8\pi}
\label{tau}
\eea

The Laplace tension in the planar case \eq{gammaplane} for $\Lambda=200$ is shown in \fig{gammag} and is superimposed with the curve of $\gamma$ for a spherical vesicle with $N=10000$. Hence for sufficiently large $N$, the analytical predictions of the planar case can be used for quasi-spherical vesicle with good confidence.

\section{Renormalisation of the surface tension}
\label{renormalisation}

The previous arguments are valid only for small height gradients or small excess areas $\alpha\ll1$ or equivalently for large surface tensions $\sigma$ (see~\fig{alphag}). This is the reason why we do not find the divergence of $\alpha$ as expected by David~\cite{membrane_book_David}, at order two in $u$. When $\sigma/(\kappa\Lambda^2)\ll1$, \eq{alphaP} is no more valid and we have to take into account the higher order terms in $(\nabla h)^4$ in the Hamiltonian.

In this section we compute the renormalisation of the surface tension $\sigma$ and the bending modulus  $\kappa$ of the Helfrich Hamiltonian in \eq{H0_pl}. We focus on planar membranes to avoid inessential complications in the spherical geometry~\cite{kleinert86}. We follow the historical Wilson's perturbative renormalisation procedure. It is worthwhile to do it carefully since different results can be found in the literature~\cite{david91,membrane_book_David,cai94}, especially for the surface tension. Moreover, we focus on the effect of the Faddeev-Popov corrective term on the renormalisation of $\sigma$ and $\kappa$, which does not lead to the same contributions according to David~\cite{david91} or Cai \textit{et al.}~\cite{cai94}. 

\subsection{Renormalisation of the Helfrich Hamiltonian}

Expanding up to order 4 the Helfrich Hamiltonian \eq{H0_pl}, one has
\bea 
 \sqrt{g} &=& \sqrt{1+(\nabla h)^2}\\
&=&1+\frac12(\nabla h)^2-\frac18(\nabla h)^4+{\cal O}(h^6)
\label{surface}
\eea
The normal to the membrane is $\bn(\bx)=(\mathbf{e}_z-\mathbf{e}_\alpha\partial_\alpha h )/\sqrt{g}$ (using $\alpha=\{x,y\}$ and the Einstein convention). By expanding the mean curvature $2H=\nabla \cdot \bn$, one finds
\be
{\cal H}_h[h(\mathbf{x})]=\sigma A_{\rm p}+ {\cal H}_0[h(\mathbf{x})]+ {\cal U}[h(\mathbf{x})]
\ee
where ${\cal H}_0[h(\bx)]$ is the Gaussian Helfrich Hamiltonian \eq{H1} and
\bea
{\cal U}[h(\mathbf{x})]&=& -\frac12 \int  \md\bx \left\{\frac{\sigma}4(\nabla h)^4 +\kappa \left[\frac12 (\nabla h)^2(\nabla^2 h)^2 \right.\right.\nonumber\\
&+& \left.  \left.  2 (\nabla^2 h)\partial_\alpha h \partial_\beta h \partial_\alpha \partial_\beta h \right]\right\}  \label{U}
\eea
Going to the Fourier space and applying Wilson's procedure to integrate over the large $q$ modes, one obtains (see~\cite{these_guillaume})
\bea
\frac14 \int  \md\bx \left\langle(\nabla h)^4\right\rangle_{h_>} &=&\int_0^{\Lambda/b} \frac{q\md q}{2\pi} q^2 |h_<(q) |^2 \nonumber\\ 
&\times&\int_{\Lambda/b}^{\Lambda} \frac{q'\md q'}{2\pi} \frac{q'^2}{\beta(\sigma q'^2+\kappa q'^4)} \label{U1f}
\eea
where $b>1$ is the scaling parameter. Using the isotropy of the membrane, one has
\be
\bigl \langle(\nabla^2 h)\partial_\alpha h \partial_\beta h \partial_\alpha \partial_\beta h\bigr\rangle_{h_>} \equiv\ \frac12\bigl \langle (\nabla h)^2(\nabla^2 h)^2 \bigr\rangle_{h_>}
\ee
which gather the two last terms of \eq{U} with a prefactor $3/2$. This final term proportional to $\kappa$ is splitted in two terms, one renormalises the bending modulus, the term in $\frac{3\kappa}2(\nabla^2 h)^2\langle (\nabla h)^2 \rangle_{h_>}$, and the second one, $\frac{3\kappa}2(\nabla h)^2\langle (\nabla^2 h)^2 \rangle_{h_>}$, renormalises the surface tension.
We finally obtain (up to an additive constant)~\cite{these_guillaume}
\be
\tilde{\cal  H}[h_<]= \frac12\int_0^{\Lambda/b} \frac{q\md q}{2\pi} \left(\tilde{\sigma} q^2+\tilde{\kappa} q^4\right)|h_<(q)|^2
+{\cal U}[h_<]
\ee
where
\bea
\tilde{\kappa}&=& \kappa-\frac{3}{2\beta}I(b,\sigma/\kappa) \label{kapparenor}\\
\tilde{\sigma}&=& \sigma\left[1+\frac{1}{2\beta\kappa}I(b,\sigma/\kappa)\right]-\frac{3}{2\beta} \int_{\Lambda/b}^{\Lambda} \frac{q\md q}{2\pi}
\label{sigmarenor}
\eea
with
\be
I(b,x)= \int_{\Lambda/b}^{\Lambda} \frac{q\md q}{2\pi}\frac1{q^2+x}=\frac1{4\pi} \ln \left(\frac{\Lambda^2+x}{\frac{\Lambda^2}{b^2}+x}\right)\label{Ib}
\ee
The last term of the rhs. \eq{sigmarenor}, which will play an important in the following, comes directly from the bending energy fluctuations. 
By renormalizing the lengths according to $\bq=\bq'/b$ and the field as $h_<(\bq)=zh'(\bq')$, one gets $\sigma'=z^2 b^{-4}\tilde{\sigma}$ and $\kappa'=z^2 b^{-6}\tilde{\kappa}$. Assuming $\kappa'=\tilde{\kappa}$, \textit{i.e.} $z=b^3$, and an infinitesimal transformation $b$ close to 1 one obtains the renormalisation equations for $\sigma$ and $\kappa$:
\bea
b\frac{\md\sigma}{\md b} &=& 2\sigma+\frac{\Lambda^2}{4\pi\beta}\left(\frac{\sigma}{\Lambda^2 \kappa +\sigma}- 3\right) \label{RGsig} \\
b\frac{\md\kappa}{\md b} &=& - \frac{3}{4\pi\beta} \frac{\Lambda^2\kappa}{\Lambda^2\kappa+\sigma} \label{RGkap}
\eea
Eq.~(\ref{RGkap}) has first been obtained by Peliti and Leibler~\cite{peliti85} and also by David and Leibler~\cite{david91} (using the effective action method~\cite{membrane_book_David}). However the equation on the surface tension $\sigma$, \eq{RGsig}, is different from the one obtained in Refs.~\cite{peliti85,meunier87,david91} where some terms have been omitted. 

The effective surface tension and bending modulus are given by $\sigma_{\rm eff}= \lim_{b\to\infty}\sigma(b)/b^2$ and $\kappa_{\rm eff} = \lim_{b\to\infty}\kappa (b)$~\cite{david91}. Since \eqs{RGsig}{RGkap} are non-linear, we look at the analytical solutions, which depends on the microscopic values $\kappa_0$ and $\sigma_0$ (the initial values for $b=1$), in two asymptotic limits. For bilayers, one has usually $\beta\kappa_0\simeq 20\gg1$ at room temperature. For very large temperatures, such that $\beta\kappa_0<1$ the preceding one loop approximation is not valid~\cite{david91}. 

In the tension regime where $\sigma\gg \kappa \Lambda^2$, \eqs{RGsig}{RGkap} simplifies to 
\be
b\frac{\md\sigma}{\md b} \simeq 2\sigma-\frac{\Lambda^2}{2\pi\beta}; \qquad
b\frac{\md\kappa}{\md b} \simeq 0 
\ee
which yields
\bea
\sigma_{\rm eff} &\simeq& \sigma_0-\frac{\Lambda^2}{4\pi\beta} \label{sigma_eff_T}\\
\kappa_{\rm eff} &\simeq& \kappa_0 
\eea
Note that contrary to David and Leibler~\cite{david91}, one obtains a (very small) renormalisation of $\sigma$. Hence as soon as $\sigma_{\rm eff} \gg \kappa_0\Lambda^2$ the membrane can be viewed as a simple fluid interface (with a very small bending energy) with a surface tension $\sigma_{\rm eff}$. 

In the opposite limit where $\sigma\ll \kappa \Lambda^2$, \eqs{RGsig}{RGkap} simplifies to
\be
b\frac{\md\sigma}{\md b} \simeq 2\sigma-\frac{3\Lambda^2}{4\pi\beta};\qquad
b\frac{\md\kappa}{\md b} \simeq -\frac{3}{4\pi\beta}
\ee
The effective elastic moduli are
\bea
\sigma_{\rm eff} &\simeq& \sigma_0-\frac{3\Lambda^2}{8\pi\beta} \label{sigma_eff_B}\\
\kappa_{\rm eff} &\simeq& \kappa_0 -\frac{3}{4\pi\beta}\ln\left(\frac{\ell}{\zeta}\right) \label{kappa_eff}
\eea
where an infrared cutoff $\ell\simeq \zeta b_{\max}$ has been introduced and $\zeta=2\pi/\Lambda$. We recover the classical result for the renormalisation of the bending rigidity~\cite{peliti85,meunier87,david91}. 

For very large membranes at fixed temperature, we can enter back to the tension regime. The critical length for which it occurs is obtained by writing $\sigma(b_c)=\kappa(b_c)\Lambda^2$, which in the limit of large $b$ leads to $\sigma_{\rm eff}=\kappa(\ell_c)/\ell_c^2$ where 
\be
\ell_c\simeq\left(\frac{\kappa_{\rm eff}^*}{\sigma_{\rm eff}}\right)^{1/2}
\ee 
with $\kappa_{\rm eff}^*\simeq \kappa_0-3\ln(\Lambda^2\kappa_0/\sigma_{\rm eff})/(8\pi\beta)$ and $\sigma_{\rm eff}$ given by \eq{sigma_eff_B}. We recover the discussion by David and Leibler~\cite{david91} where for $\ell>\ell_c$ the surface tension $\sigma_{\rm eff}$ dominates whereas it is negligible for $\ell<\ell_c$. The critical length $\ell_c$ is therefore the renormalized correlation length, which was $\xi$ at the Gaussian order, and we have checked that $\ell_c\gtrsim\xi$ for $\kappa_0\Lambda^2\gtrsim\sigma_0$ and $\ell_c\gg\xi$ for $\kappa_0\Lambda^2\gg\sigma_0$.

For $\sigma_0= 3\Lambda^2k_{\rm B}T/(8\pi)$, one has a membrane with zero effective surface tension ($\ell_c\to\infty$) and which is only controlled by the bending rigidity. However, for $\ell>\xi_p$, where 
\be
\xi_p\equiv \zeta\exp\left(\frac{4\pi\beta\kappa}3\right)\label{PGG}
\ee
is the de Gennes-Taupin persistence length (defined by $\kappa_{\rm eff}(\ell=\xi_p)=0$)~\cite{PGG82}, the membrane is crumpled, \textit{i.e.} a purely entropic membrane of independent stiff patches of typical size $\xi_p$, the 2D equivalent of a freely jointed polymer. For a lipid bilayer at room temperature $\beta\kappa>10$, which leads to $\xi_p>\zeta e^{40}$. This crumpled membrane might be encountered for monolayers in the sponge phase~\cite{PGG82} but not for vesicles.

\begin{figure}
\centering
\includegraphics[width=0.99\linewidth]{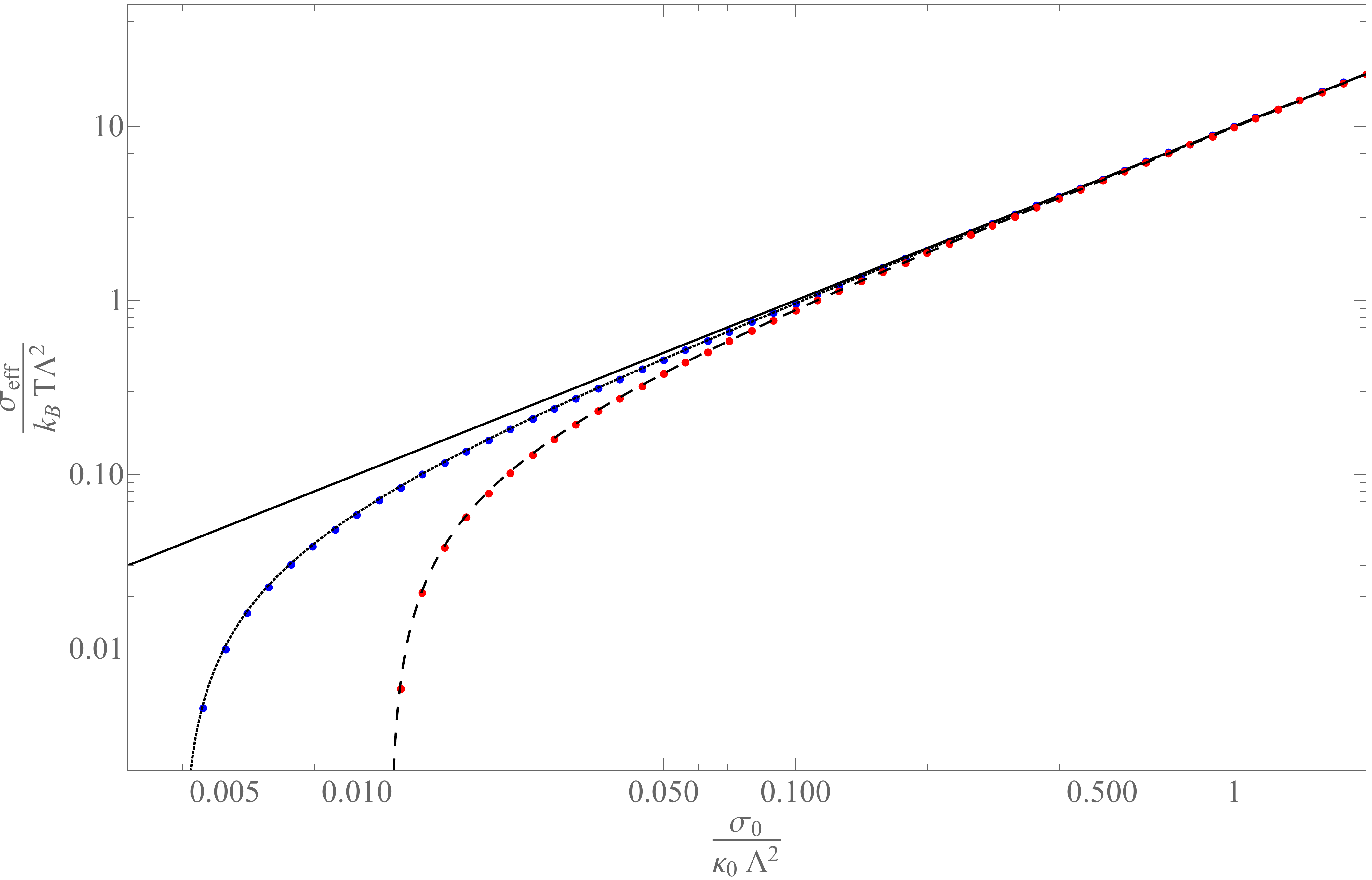} 
\caption{Renormalised surface tension $\sigma_{\rm eff}$ vs. $\sigma_0/(\kappa_0 \Lambda^2)$, solution of \eqs{RGsig}{RGkap}  (in red) and \eqs{RGsigFP}{RGkap} (in blue) for $\beta\kappa_0=10$. Are also shown the $\sigma_{\rm eff}=\sigma_0$ (straight solid line) and the asymptotic formulas \eqs{sigma_eff_B}{sigma_eff_B_FP} for $\sigma_0\ll\kappa_0 \Lambda^2$ (log-log plot).\label{renorm}}
\end{figure}
The numerical solution $\sigma_{\rm eff}$ is plotted against $\sigma_0/\kappa_0$ in \fig{renorm} (red dots). The agreement with the asymptotic solutions \eqs{sigma_eff_T}{sigma_eff_B} is very good, even in the intermediate range.

\subsection{Modification of the renormalisation equations by the Faddeev-Popov term}
\label{FP}

One important issue of the preceding calculation concerns the degrees of freedom on which the functional integral is performed. An integral over all the degrees of freedom $\int \mathcal{D}\bX$ unnecessarily counts many configurations $\bX(\bu)$ which are equivalent under some change of parametrisation and therefore describe the same membrane. To avoid this over-counting, the measure must be weighted by a correction factor, $\int \mathcal{D}\bX J[\bX]\delta[f_i(\bX)]$ where $J[\bX]$ is called the Faddeev-Popov determinant and $f_i(\bX)=0$ are two constraints that fix the coordinate choice. This amounts to fix a gauge as well explained by David in Ref.~\cite{membrane_book_David}, and the only configurations entering in the integral are now those obeying this gauge condition.

If one chooses the normal gauge with a reference background configuration, noted as $\bX_0(\bu)$, these constraints ensure that the new configurations $\bX(\bu)$ are such that $ \bX(\bu)=\bX_0(\bu) + \bR(\bu)$ with $f_i(\bX)\equiv\bR(\bu)\cdot\partial_i\bX_0(\bu)=0$. 
In the literature, two different results have been obtained for the Faddeev-Popov determinant $J$ in the normal gauge. On the one hand, David~\cite{membrane_book_David} finds
\be
J[\bX]=\prod_\bu \det[g_{ij,0}+\partial_j\bR(\bu)\cdot\partial_i\bX_0(\bu)]
\label{JD}
\ee
where $g^0_{ij}$ is the metric of the background configuration $\bX_0(\bu)$. For the Monge gauge, $\bX_0(\bu)$ defines a plane parametrised by a Cartesian system of coordinates, $\bX_0(x,y)=x\mathbf{e}_x+y\mathbf{e}_y$. The metric simplifies to $g^0_{ij}=\delta_{ij}$ and $\partial_j\br(\bx)\cdot\partial_i\bX_0(\bx)=0$, and \eq{JD} leads to $J=1$.

On the other hand, Cai \textit{et al.}~\cite{cai94} obtained a different formula for $J$:
\be
J[\bX]=\prod_\bu \frac{\det[g_{ij,0}+\partial_j\bR(\bu)\cdot\partial_i\bX_0(\bu)]}{\sqrt{g}}
\label{JC}
\ee
where $g=\det g_{i,j}$ is the determinant of the metric associated to the configuration $\bX(\bu)$. The only difference with \eq{JD} is the $g^{-1/2}$ factor. It comes from the fact that when doing the infinitesimal re-parametrisation from $\bX$ to $\bX'$ necessary to compute $J$, the area per patch $a=A/N$ is kept constant for all the patches, whereas in David's derivation it is $a_{\rm p}=A_{\rm p}/N$ which is kept constant. It also means that, in the Monge gauge, this infinitesimal displacement is done along the normal $\bn$ and not along $\mathbf{e}_z$.
Since only displacements along $\mathbf{e}_z$ are taken into account ({\it i.e.} integration over $h(x,y)$) in the integral, a shorter displacement, corrected by the factor $\delta h\  \bn\cdot\mathbf{e}_z=\delta h(\bx)/\sqrt{g(\bx)}$ comes out, corresponding to a virtual displacement $\delta h$ along $\bn$.
Although the difference between the two approaches is subtle, it has important consequences. Therefore in the Monte Carlo simulations presented in the next Section, we will consider the two types of displacement. Note that for a quasi-spherical vesicle the background configuration is the sphere having the same volume of radius $R$. Therefore we have the following correspondence: $A_{\rm p}\to A_{\rm s}=4\pi R^2$ and $\mathbf{e}_z\to \mathbf{e}_r$.

In Cai \textit{et al.} formulation, the measure is changed according to
\bea
\int J\mathcal{D}h &=& \int \mathcal{D}h \prod_\bx \frac1{\sqrt{g(\bx)}}=\int \mathcal{D}h e^{-\frac1{2\beta}\int \md\bx\mathrm{tr}\ln g}\nonumber\\
&\simeq&\int \mathcal{D}h \ e^{-\frac1{2\beta a^2} \int \md\bx (\nabla h)^2}
\eea
The term in the last exponential must be added to the perturbation term $ {\cal U}[h(\mathbf{x})]$ (since it is on the order of $\beta^{-1}$) and only modifies the renormalisation of the surface tension by adding a term in $1/\zeta^2=\int_{\Lambda/b}^\Lambda q\md q/(2\pi)$ to \eq{sigmarenor}. It modifies \eq{RGsig} according to
\be
b\frac{\md\sigma}{\md b} = 2\sigma+\frac{\Lambda^2}{4\pi\beta}\left(\frac{\sigma}{\Lambda^2 \kappa +\sigma}- 1\right) \label{RGsigFP}
\ee
\textit{i.e.} the 3 in the brackets of the rhs. of \eq{sigmarenor} has been changed into 1. 

It modifies the result for the effective surface tension. In the tension regime, $\sigma_0\gg\kappa_0\Lambda^2$ one finds now no renormalisation of the elastic parameters by the thermal fluctuations, $\sigma_{\rm eff}=\sigma_0$ and $\kappa_{\rm eff}=\kappa_0$. This is an interesting result: the Faddeev-Popov term exactly cancels (at order $\beta^{-1}$) the previous renormalisation of $\sigma_0$, that is it somehow compensates the term given in \eq{U1f} to keep $A$ constant.

In the bending regime, one obtains
\bea
\sigma_{\rm eff} &\simeq& \sigma_0-\frac{\Lambda^2}{8\pi\beta} \label{sigma_eff_B_FP}\\
\kappa_{\rm eff} &\simeq& \kappa_0 -\frac{3}{4\pi\beta}\ln\left(\frac{\ell}{\zeta}\right)
\eea
The very good agreement between the full numerical solution (blue dots) and the asymptotic formulae is shown in \fig{renorm}.
The interpretation in this regime is more involved. Geometrically, the introduction of the Faddeev-Popov correction term [\eq{JC}] clearly corresponds to smaller moves along $\mathbf{e}_z$, which globally leads to a less fluctuating membrane area ${\cal A}$. It does not modify $\kappa_{\rm eff}$ at this order but $\sigma_{\rm eff}$ is less reduced (by a factor $\Lambda^2/(8\pi\beta)$) by thermal fluctuations.

In the following, we note the renormalized surface tension
\be
\sigma_{\rm eff} \simeq \sigma-\epsilon\frac{k_{\rm B}T\Lambda^2}{8\pi} \label{sigma_eff_eps}
\ee
where $\epsilon=1$ (respectively $\epsilon=3$) when the Faddeev-Popov correction is taken according to Cai \textit{et al.}, \eq{sigma_eff_B_FP}, (respect. David, \eq{sigma_eff_B}). The value $\sigma_c=\epsilon k_{\rm B}T\Lambda^2/(8\pi)$ for which $\sigma_{\rm eff}$ vanishes is called the residual tension.

\subsection{Renormalised Laplace and frame tensions}

How is the excess area changed when the renormalized surface tension $\sigma_{\rm eff}$ is taken into account? 
One way to obtain an approximate formula is to keep the truncature at order 2 in the Hamiltonian by replacing the $\sigma$ and $\kappa$ appearing in the second term of the rhs. of \eq{H0_pl} by the renormalized values $\sigma_{\rm eff}$  and $\kappa_{\rm eff}$ given in \eqs{kappa_eff}{sigma_eff_eps}. Hence the excess area \eq{alphaP} becomes
\be
\alpha_{\rm pl,eff} = \frac{k_{\rm B}T}{8\pi \kappa_{\rm eff}} \ln\left(1+\frac{\kappa_{\rm eff}\Lambda^2}{\sigma_{\rm eff}}\right) 
\label{alpha_eff}
\ee
where we assume $\kappa_{\rm eff}>0$, that is the typical length scale $L$ of the membrane remains smaller than the de Gennes-Taupin  persistence length, \eq{PGG}.
In the limit $\sigma\to\sigma_c$, the renormalized surface tension vanishes and one has to carefully take into account the infrared cutoff
\be
\alpha_{\rm pl,eff}(\sigma_{\rm eff}\to0)=\frac{k_{\rm B}T}{4\pi \kappa_{\rm eff}} \ln\left(\frac{L\Lambda}{2\pi}\right)
\label{alpha_eff_cutoff}
\ee
and $\alpha_{\rm pl,eff}$ increases logarithmically with the size of the system and diverges when $L\to\infty$.
Using a critical study of membranes under tension in the limit $D\to\infty$ (where $D$ is the dimension of space), David and Guitter show in \cite{david86,david87} that when $\sigma_{\rm eff}\to0$ a phase transition to a crumpled membrane occurs where the total area of the membrane diverges.

Repeating this procedure for the calculation of the frame surface tension $\tau$ given by \eq{tau} leads to
\bea
\tau_{\rm eff} &=& \sigma + \sigma_{\rm eff}\alpha_{\rm pl,eff} - \frac{k_{\rm B}T\Lambda^2}{8\pi}\label{taur1}\\
&=& \sigma_{\rm eff}(1+\alpha_{\rm pl,eff}) + (\epsilon-1)\frac{k_{\rm B}T\Lambda^2}{8\pi}
\label{taur}
\eea
The renormalized Laplace tension becomes
\be
\gamma_{\rm eff} = \sigma_{\rm eff}+ (\epsilon-1)\frac{k_{\rm B}T\Lambda^2}{8\pi(1+\alpha_{\rm pl,eff})}
\label{gamr}
\ee
In both cases, $\tau_{\rm eff}$ and $\gamma_{\rm eff}$ are always positive for $\sigma_{\rm eff}\geq0$, and $\gamma_{\rm eff} \to0$ when $\sigma_{\rm eff}\to0$ (for $L\to\infty$).  
Interestingly for the case $\epsilon=1$, it simplifies to $\tau_{\rm eff} =\sigma_{\rm eff}(1+\alpha_{\rm pl,eff})$ and $\gamma_{\rm eff} =\sigma_{\rm eff}$ which implies furthermore that $\tau_{\rm eff} \to0$ for $\sigma_{\rm eff}\to0$. 

Within this framework, the transition when $\sigma\to\sigma_c$ corresponds to a membrane with a very large area $A$ and where $\tau_{\rm eff}\simeq 0$, which provides a way to characterise the transition experimentally or numerically (see below).

\section{Numerical methods}
\label{Num_Meth}

In this Section we describe the algorithm used to simulate the vesicles. First we describe the discretisation of the Helfrich Hamiltonian in \eq{H0} and the implementation of the volume and area constraints. Then we describe the Monte Carlo (MC) processes and specifically the two different types of MC moves we implemented.

\subsection{Tessellation of the initial sphere}

The first step consists in creating the discrete surface of a sphere by using a simple rule to keep track of the connectivity of all the elementary surface patches of the system. It will be helpful  to compute the local principal curvatures.

We begin with a regular icosahedron, composed by 20 triangular faces named $I_i$ with $i=\{1,2,\dots,20\}$, and each of the 12 vertex belongs to a sphere of radius $R$ with spherical coordinates $\bx_1=(R,0,0)$, $\bx_j=(R,\pi/2-\arctan(1/2),2\pi(j-1)/5)$ for $j=\{2,\dots,6\}$, $\bx_j=(R,\pi/2+\arctan(1/2),2\pi(j-7)/5)$ for $j=\{7,\dots,11\}$, and $\bx_{12}=(R,\pi,0)$. 

\begin{figure}
\centering
\includegraphics[width=\columnwidth]{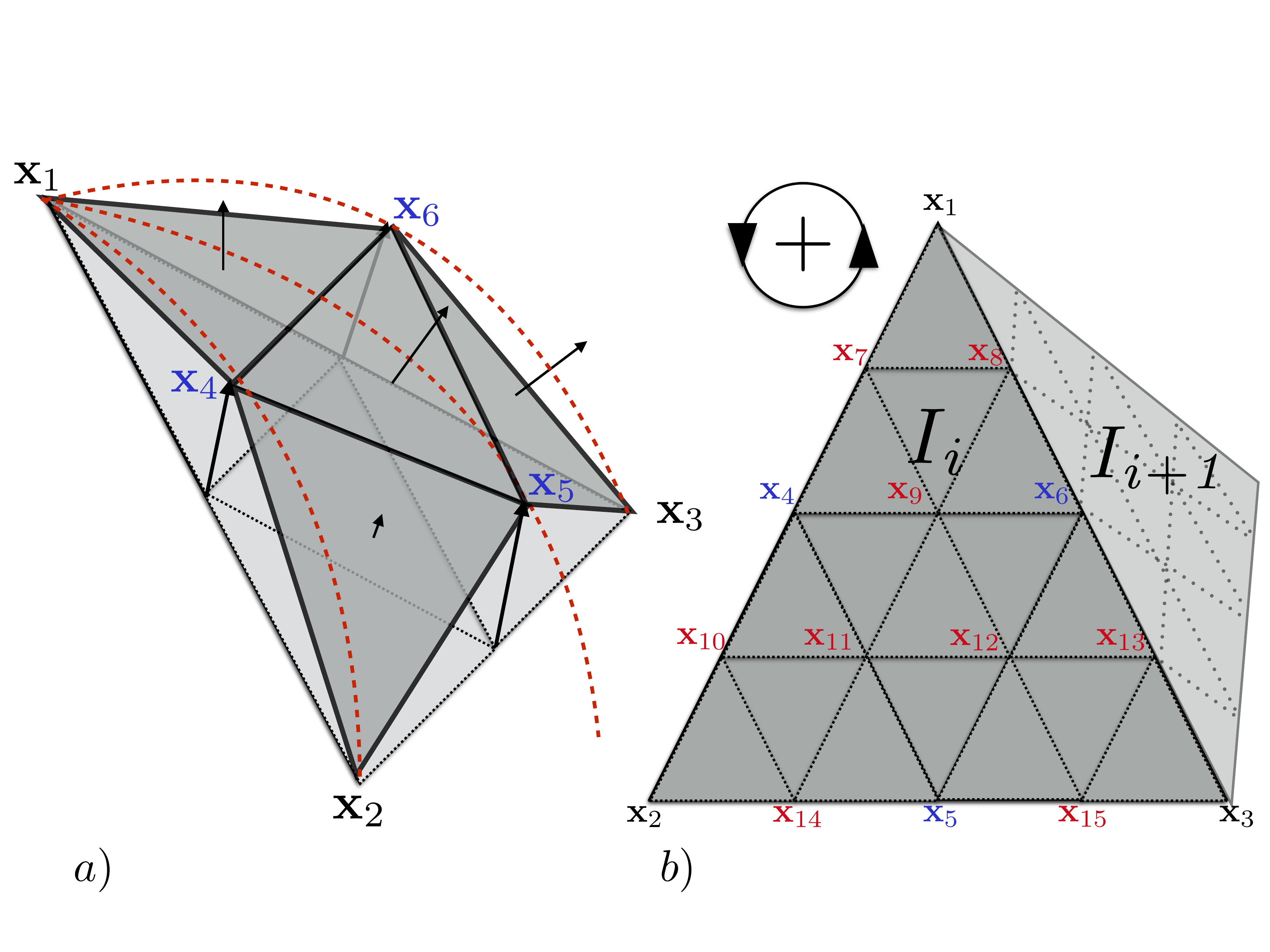}
\caption{\label{sketchvoisin} Tessellation of the sphere. (a) Sketch illustrating the creation of first generation of vertices (in blue) from the initial triangle (vertices in black). The red lines corresponds to great circles of the sphere of radius $R$. (b) Projection on a plane of the initial triangle $I_i$ with the new created small triangles (the red vertices correspond to second generation) showing the connectivity between the triangles.}
\end{figure}

To make the sphere tessellation, we choose one triangle with vertices $\bx_ j$ (with $j=1,2,3$) (Cf. fig.~\ref{sketchvoisin}.a). By taking the middle of each edge of the triangle and projecting it on the sphere of radius $R$ we get 3 new vertices $\bx_j$, and we create 4 triangles from one. If $n$ is the number of iterations of the process, we eventually obtain $N=10\times 4^n+2$ points.
 
To have a smart storage process which allows us to keep in memory the connectivity of all the points, we associate the array $T_i[l,m]$ of size $3(2n+1)^2$ with each of the 20 initial triangles $I_i$, in which are stored the positions of all the vertices $\bx_j$ after the subdivisions. We fill the $T_i$ as follows~:
\begin{center}
$\begin{pmatrix}
\bx_1&0&0&0&0\\
\bx_2&\bx_3&0&0&0\\
0&0&0&0&0\\
0&0&0&0&0\\
0&0&0&0&0&
\end{pmatrix} 
\rightarrow\begin{pmatrix}
\bx_1&0&0&0&0\\
\bx_4&\bx_6&0&0&0\\
\bx_2&\bx_5&\bx_3&0&0\\
0&0&0&0&0\\
0&0&0&0&0&
\end{pmatrix} 
 \rightarrow\begin{pmatrix}
\bx_1&0&0&0&0\\
\bx_7&\bx_8&0&0&0\\
\bx_4&\bx_9&\bx_6&0&0\\
\bx_{10}&\bx_{11}&\bx_{12}&\bx_{13}&0\\
\bx_2&\bx_{14}&\bx_5&\bx_{15}&\bx_3&
\end{pmatrix} \rightarrow\ \cdots$
\end{center}
We thus keep track of the ``natural'' order of the system (see \fig{sketchvoisin}b). For the vertices which are not on the edges, located at position $(l,m)$ in the array, the coordination number is 6 and the neighbours are simply located at $(l-1,m)$, $(l-1,m-1)$, $(l,m-1)$, $(l+1,m-1)$, $(l+1,m)$, $(l+1,m+1)$. Finally two matrices are constructed to store only once the positions of the vertices and the connectivity between them. Note that the 12 initial vertices have a coordination number 5, but one can assume that for $N \sim 10^3$, the consequences of this lower coordination are negligible.

\subsection{Discretised Hamiltonian and geometrical quantities}

\begin{figure}
\centering
\includegraphics[width=\columnwidth]{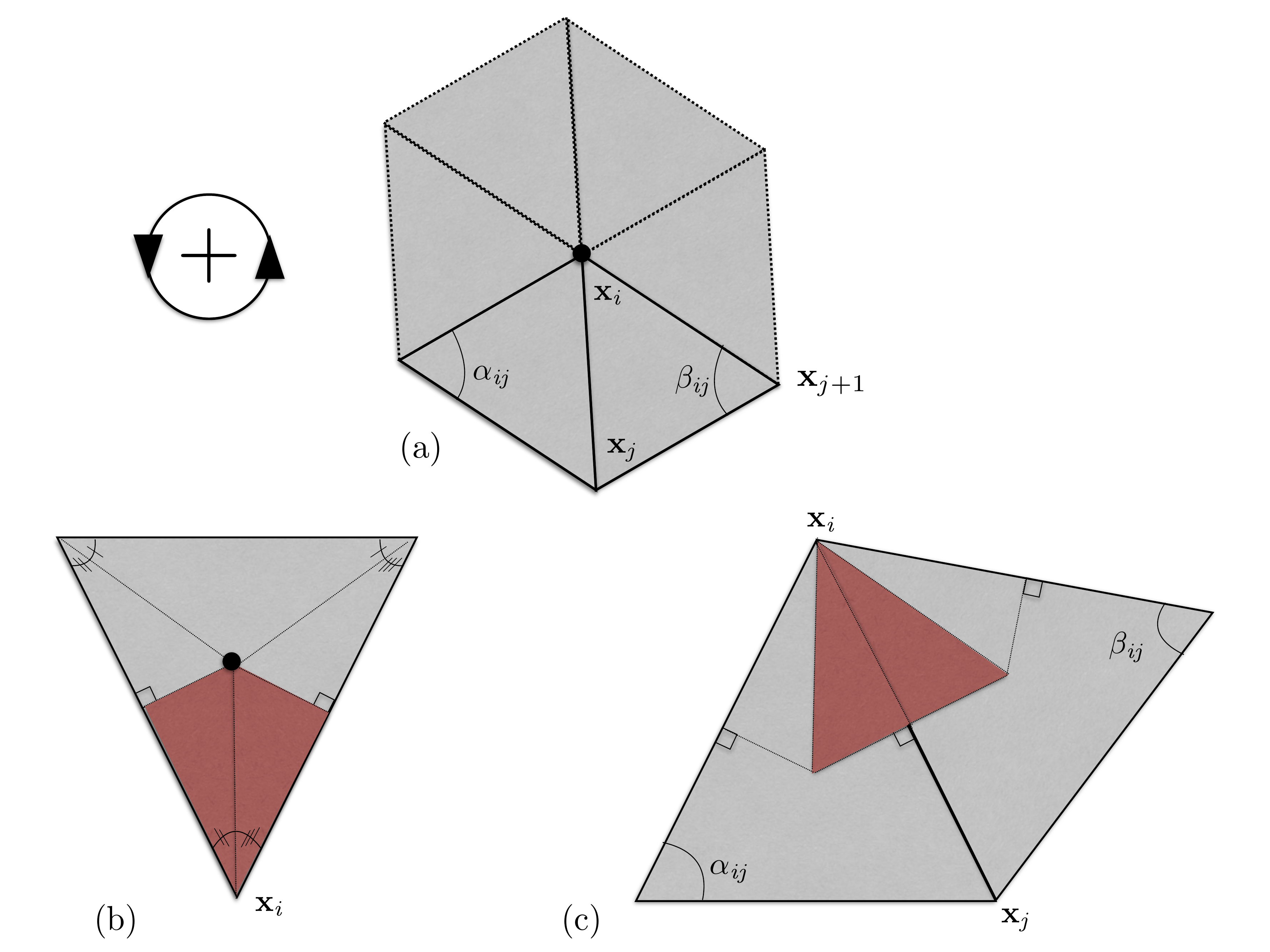}
\caption{\label{sketch} (a)~First neighbours of the vertex $\bx_i$ with a connectivity 6 with the definition of the angles $\alpha_{ij}$ and $\beta_{ij}$ used to compute the Laplace-Beltrami operator~\eq{laplace-beltrami}. (b) Geometrical definition of the Vorono\"{i} area of a triangle associated with the vertex $\bx_i$ [see \eq{aireT}]. (c)~Averaged Vorono\"{i} area ${\cal A}_{ij}$ appearing in \eq{airemixed}.}
\end{figure}

To compute the vesicle curvature, volume and area in one step, we use the method of Meyer \textit{et al.}~\cite{meyer2003}. It is exact as compared to approximate methods commonly used to simulate vesicles~\cite{sunilkumar1997,hu2011,amazon2013,amazon2014}.
First we need a discretised version of the Helfrich Hamiltonian, \eq{H0}~:
\begin{equation}
{\cal H}_h =  \frac12\sum_i   \ \kappa(2H_i - C)^2 \ {\cal A}_i
\label{H0D}
\end{equation}
where $|2H_i|$ is the norm of the discrete Laplace-Beltrami operator for a triangulated surface~\cite{meyer2003}:
\begin{equation}
{\bold K}_i=  \frac1{2{\cal A}_i} \sum_j  \left( \cot\alpha_{ij}+\cot\beta_{ij} \right) (\bx_i-\bx_j)
\label{laplace-beltrami}
\end{equation}
$\bx_i$ is the position of vertex $i$, $\alpha_{ij}$ and $\beta_{ij}$ are the angles of the two triangles sharing the edge $\bx_i\bx_j$ and opposite to this edge (\fig{sketch}a). The sum is taken over the neighbours of $i$. 
The sign of $2H_i$ is given by the scalar product of the outgoing normal vector to the surface ${\bold n}_i={\bold K}_i/||{\bold K}_i||$ with $\bx_i$.

The area $ {\cal A}_i$ is the Vorono\"{i} area:
\be
{\cal A}_i= \sum_j{\cal A}_{ij}=\frac18 \sum_j  \left( \cot\alpha_{ij}+\cot\beta_{ij} \right) || \bx_i-\bx_j||^2
\label{airemixed}
\ee
where ${\cal A}_{ij}$ is represented in \fig{sketch}c and the sum is over the edges connected to $\bx_i$. The area of the triangle of \fig{sketch}b is simply:
\bea
{\cal A}_i = \sum_j{\cal A}'_{ij} &=& \frac18 \sum_j  \left( \cot\alpha_{ij+1} || \bx_i-\bx_{j+1}||^2\right.\nonumber\\
&+& \left.\cot\beta_{ij} || \bx_i-\bx_j||^2\right)
\label{aireT}
\eea
If the triangle $\Delta$ has an obtuse angle, one has to modify the definition of the  Vorono\"{i} area: either it is obtuse at $\bx_i$ and ${\cal A}'_{ij}=  {\rm area}(\Delta)/2$, or  ${\cal A}'_{ij}= {\rm area}(\Delta)/4$~\cite{meyer2003}.

The volume related to a vertex $\bx_i$ is the sum of the volumes of the adjacent tetrahedrons of base the red area of \fig{sketch}c and of summit the center of the vesicle:
\be
{\cal V}_i= \sum_j{\cal V}_{ij}=\frac13\sum_j {\cal A}'_{ij} \ \bx_i\cdot{\bold n}_{f_j} 
\ee
where ${\bold n}_{f_j}$ is the normal vector of the face $j$: 
\be
{\bold n}_{f_j}=\frac{(\bx_j-\bx_i)\times(\bx_{j+1}-\bx_j)}{||(\bx_j-\bx_i)\times(\bx_{j+1}-\bx_j)||}
\ee

Depending on the thermodynamical ensemble we are working with, we impose numerically one or two constraints on the total volume ${\cal V}=\sum_i{\cal V}_i$ and the area $\mathcal{A}=\sum_i{\cal A}_i $. In the $(A, V)$ ensemble, we impose 2 hard constraints:
\be
{\cal H}_c=  \frac{K_A}2 \left(\frac{\mathcal{A}}{A}-1\right)^2 \  +\frac{K_V}{2}\left(\frac{\mathcal{V}}{V}-1\right)^2  
\label{IF}
\ee
where the coefficients $K_A,K_V \sim 10^{4-6}k_{\rm B}T\gg k_{\rm B}T$. The area and the volume can still fluctuate a little, which is necessary to enable the Monte Carlo moves ($\Delta \mathcal{A}/\mathcal{A}\sim k_{\rm B}T/K_A$, $\Delta \mathcal{V}/\mathcal{V}\sim k_{\rm B}T/K_V$). The reference volume is  $V=4/3\pi R^3$  and for the area, $A=4 \pi R^2 (1+\alpha)$ where $\alpha>0$ is the excess area. The $(\sigma, V)$ ensemble is the most popular one because keeping fixed $\mathcal{A}$ and $\mathcal{V}$ at the same time using \eq{IF} proves to require prohibitively small MC steps. We rather use
\be
{\cal H}_c=  \sigma \mathcal{A} \  +\frac{K_V}{2}\left(\frac{\mathcal{V}}{V}-1\right)^2 
\label{OF}
\ee 
where $\sigma$ is the bare surface tension. An important difference with some previous studies~\cite{hu2011,amazon2013} is that  we do not impose any hardcore potential between the vertices, so that the length of the triangle edges is not enforced locally to a prescribed value but is free to change as soon as the global constraints \eqs{IF}{OF} are fulfilled. In Refs.~\cite{hu2011,amazon2013,penic} local constraints were added on the bond lengths which stabilise the structure. However, it prevents large deformations away from the sphere and makes the surface tension difficult to control. Note that since we do not impose such local constraints, a vertex is not associated to a specific group of lipid and we do not need to apply bond flips in the MC moves.

To avoid the vesicle isotropy breaking, we also apply a restoring force on the vesicle centre of position $\bx_{O'}=\sum_i \bx_i{\cal A}_i/{\cal A}$  deriving from the potential
\be
{\cal H}_O= \frac{K_O}{2R^2}||\bx_{O'}||^2  \label{OO}
\ee
where $K_O\sim 10^3-10^4\ k_{\rm B}T$. This prevents global translations of the vesicle.

\subsection{Metropolis algorithm}

We use the classical Metropolis algorithm. To avoid eventual mix-up in vertex ordering and to compare our numerical model with analytical models (see below), we choose radial MC moves without any orthoradial component. At each time step of duration $\tau_{\rm MC}$, we radially move one randomly chosen vertex, $\bx_i(t+\tau_{\rm MC})=\bx_i(t)+ d\bx_i$  where the radial vector $d\bx_i$ is defined by
\be
d\bx_i=\delta_r\ \frac{\bx_i-\bx_{O'}}{||\bx_i-\bx_{O'}||}  =\delta _r {\bold e}_r
\label{radial_dx}
\ee
The sign of $\delta _r$ is random, whereas its norm is chosen such that the acceptance ratio is about $80\%$. As explained in Section~\ref{FP} it corresponds to the choice of David for the Faddeev-Popov determinant, $J=1$.

We shall also consider another type of radial moves so that their projection onto the normal vector ${\bold n}_i$ (defined in \eq{laplace-beltrami}) is kept constant, corresponding to the Cai \textit{et al.} choice:
\be
d\bx_i^{\rm FP}=\frac{\delta _r {\bold e}_r}{{\bold n}_i\cdot{\bold e}_r}
\label{FP_dx}
\ee
where FP stands for Faddeev-Popov~\cite{cai94}, since it is a way to take into account this corrective term~\cite{cai94,shiba}, at least in an approximate way.
Indeed as already explained by Shiba \textit{et al.}~\cite{shiba}, taking these corrected FP MC moves \eq{FP_dx} amounts to modify the measure according to Cai \textit{et al.}'s choice: 
\be
\frac{\delta h_i}{|d\bx_i^{\rm FP}|}=\frac{\delta h_i}{\delta _r}\bn_i\cdot\mathbf{e}_r=\frac{\delta h_i}{\sqrt{g(\bx_i)}}
\ee
Note that in \cite{shiba} it was exponentiated in the effective Hamiltonian, with the following corrective term to $\mathcal{H}_{\rm h}$
\be
\mathcal{H}_{\rm c}=-k_{\rm B}T\sum_i \ln\cos\theta_i
\label{shiba_corr}
\ee
where $\cos\theta_i={\bold n}_i\cdot \mathbf{e}_z$.  Alternatively, in our simulations we directly modify the MC moves.

\subsection{Correlation times and statistical errors}

To measure the correlation times of the vesicle, we compute the time auto-correlation function of the positions $u$:
\be
\chi(s)=\frac{\sum_{i=1}^{n}\sum_{\alpha=1}^{\alpha_{\max}-s} v_i (t_\alpha)  v_i (t_{\alpha+s}) \sqrt{{\cal A}_i (t_\alpha)} \sqrt{{\cal A}_i (t_{\alpha+s})} }{\sum_{i=1}^{n}\sum_{\alpha=1}^{\alpha_{\max}-s} \sqrt{ {\cal A}_i (t_\alpha) } \sqrt{{\cal A}_i (t_{\alpha+s})} }\label{autocorrel}
\ee
Here $v_i (t_\alpha)=r_i (t_\alpha)-R(t_\alpha)$ where $r_i=||\bx_i-\bx_{O'}||$, and  $R(t_\alpha)=\sum_i r_i {\cal A}_i/{\cal A}$. At long times we measure $\chi(t)\sim \exp(-t/\tau_{\rm corr})$, which defines the (longest) correlation time $\tau_{\rm corr}$. These times are increasing with $N$, from $\tau_{\rm corr}(N=162)\sim10^4$~MC steps to  $\tau_{\rm corr}(N=2562)\sim10^8$~MC steps. The next possible value for $N$ is 10242 which leads to a much larger $\tau_{\rm corr}\sim10^{10}$~MC steps and therefore poor statistics. In the following, we limit ourselves to $N=642$ and 2562.

To estimate our statistical errors associated with the time averages, we compute the standard deviation for the measure of any observable $m$, which takes into account the number of independent measures, following:
\be
\Sigma(m)=\sqrt{\frac{2 \tau_{\rm corr}}{t_{\max}}(\langle m^2 \rangle -{\langle m\rangle}^2)}
\ee
where $t_{\max} \sim10^9$ to $10^{10}$~MC steps is the duration of the simulation. This formula is valid provided that $t_{\max} \gg\tau_{\rm corr}$ which is always the case in our simulations (error bars in the forthcoming plots correspond to $2\Sigma$). The statistical errors related to the determination of any fitting parameter, such as the fluctuation tension $r$, are computed using the bootstrap method~\cite{efron}.

\section{Numerical results on vesicles}
\label{Num_Res}

\subsection{Bending energy and roughness}

\begin{figure}[t]
\begin{center}
(a)\includegraphics[width=0.9\linewidth]{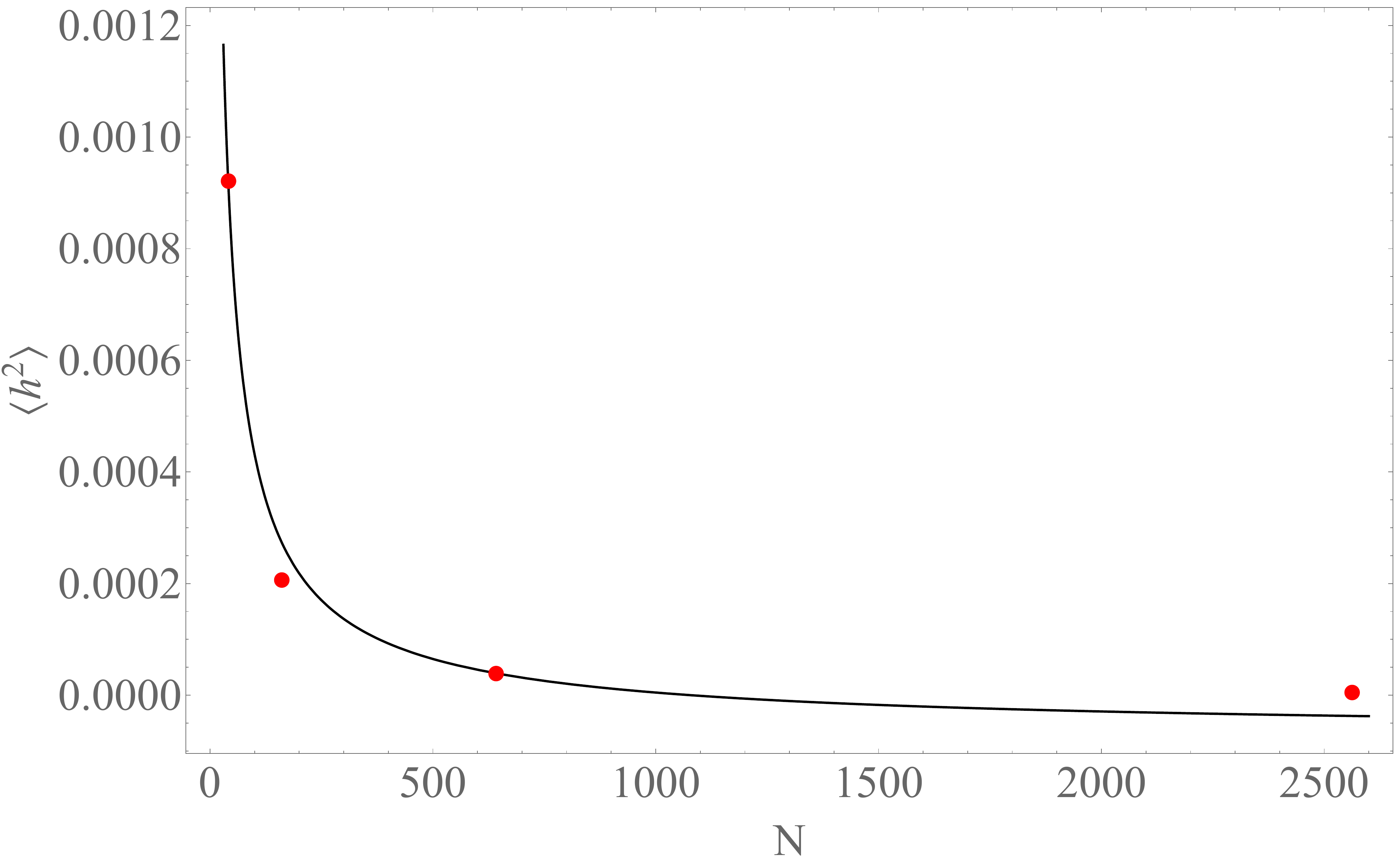}
(b)\includegraphics[width=0.9\linewidth]{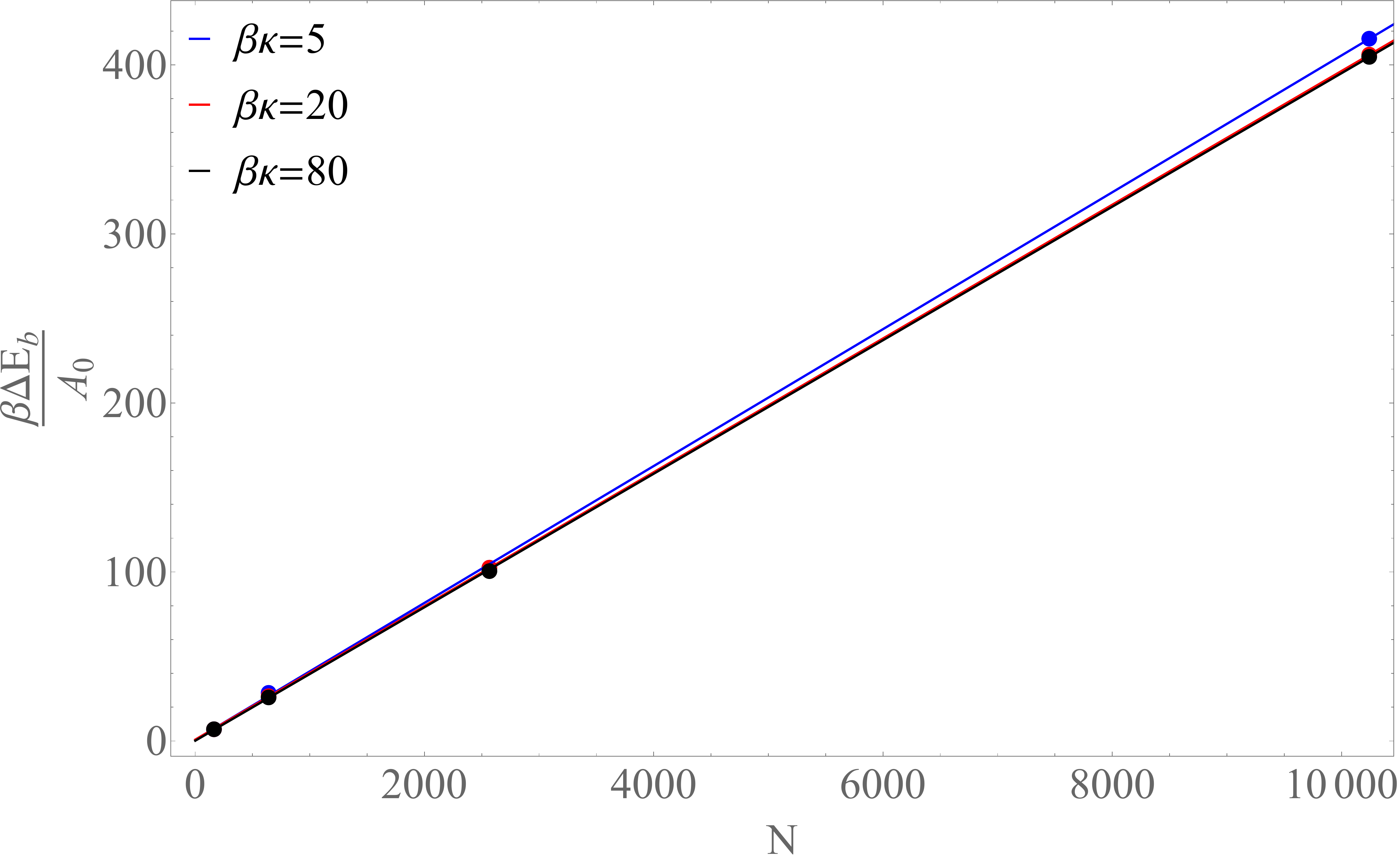}
\end{center}
\caption{(a) Roughness $\langle h^2\rangle=R^2\langle u^2\rangle$ as a function of the number of vertices $N$ for various bending moduli $\kappa$ (the spontaneous curvature fixed to $C=2/R$ where $R$ is given by the fixed volume $V$), and (b) average curvature energy $\beta\Delta E_{\rm b}/A$. Points corresponds to simulation results and lines are linear fits. Parameter values are fixed to $\alpha = 0.005$, $R=1$, $\beta K_V=\beta K_A=10^6$, $t_{\rm max}=10^9$ and (a)~$\beta\kappa=25$ (b)~$\beta\kappa=5,20,80$.\label{curv}}
\end{figure}

First to check the equilibrium properties of the simulated vesicle, we measure its roughness in the $(A,V)$ ensemble by varying $N$ and ensuring that $\alpha$ stays close to $0.5\%$. It allows a comparison to the theoretical result obtained by Seifert~\cite{seifert1995} for the roughness in this ensemble:
\be
\langle u^2\rangle = \frac{2\alpha}{N} \sum_{l\geq2}^{l_{\max}} \frac{(2 l+1)}{l(l+1)-2}  \simeq 2 \alpha  \frac{\ln N}{N} 
\label{seif}
\ee
The measured roughness is plotted in~\fig{curv}(a) as a function of $N$. The dependence in $\ln N/N$ is well verified in our simulations. The prefactor predicted by Seifert is $2\alpha=0.010$ in very good agreement with the fitted one equal to 0.011.  

Similarly, computing geometrically the curvature of the vesicle, $2H$, by using the Laplace-Beltrami operator, we plotted in \fig{curv}(b) the average dimensionless bending energy per unit surface $\beta\Delta E_{\rm b}/A$ [defined in \eq{H0}] as a function of $N$ between 162 and 10242 for $\alpha=0.005$ and $\beta\kappa=5,20$ and 80. We first observe that this bending energy is essentially independent of $\beta\kappa$ and increases linearly with $N$, which is an illustration of the classical equipartition theorem. More quantitatively we fitted the data by a linear law and found a slope equal to 0.04. This result is in excellent agreement with \eq{Hcurv} which can be rewritten as
\be
\beta\frac{\Delta E_{\rm b}}{A}\simeq  \frac{N}{2A} +\mathcal{O}(\ln N)
\label{Hh_num}
\ee 
with a theoretical slope $1/(2A)=1/[8\pi(1+\alpha)]=0.0396$ (since $A_{\rm p}=4\pi$ for $R=1$).
These two plots confirm that, for these parameter values, the simulated vesicle is in thermodynamical equilibrium. 
We have checked (not shown) that the two ensembles $(A,V)$ and $(\sigma,V)$ are equivalent.

\begin{figure}[t]
\begin{center}
\includegraphics[width=0.9\linewidth]{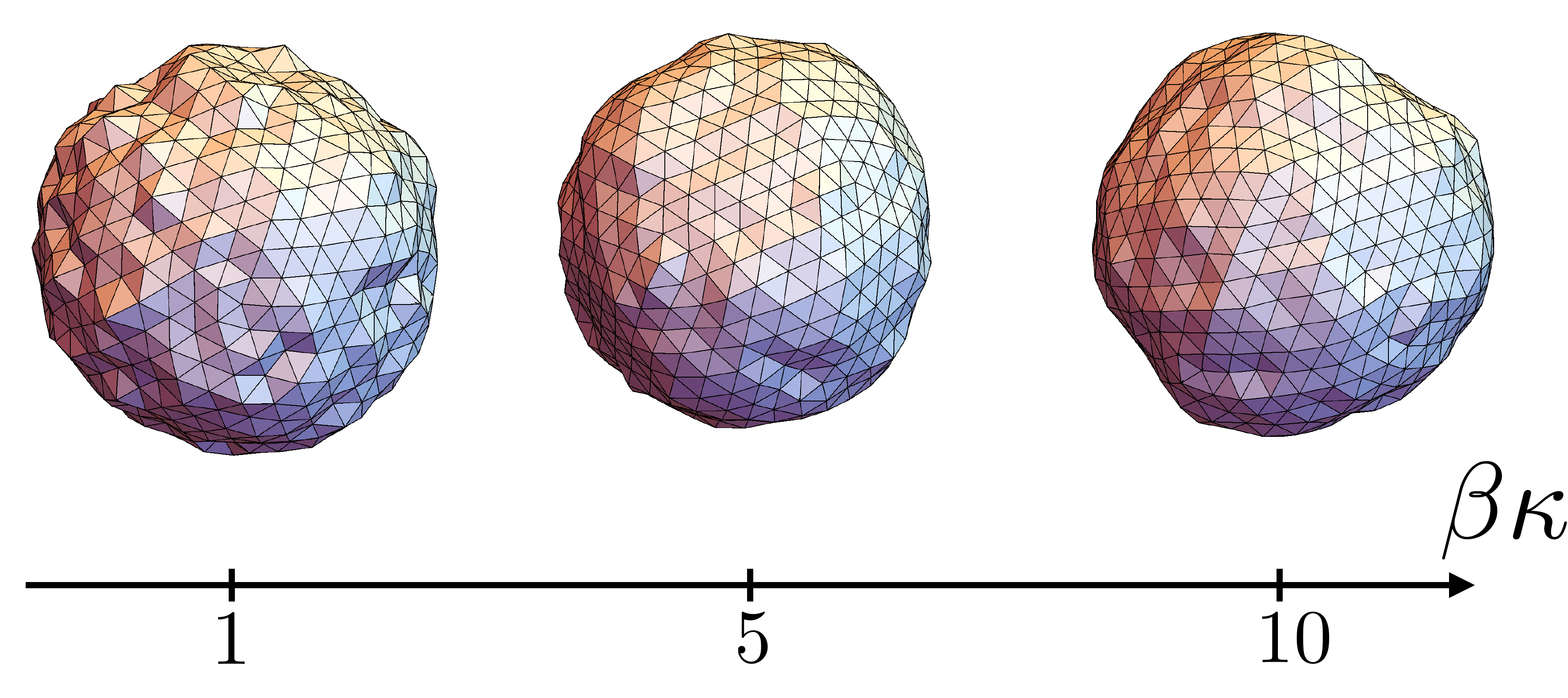}
\end{center}
\caption{Snapshots of the vesicle for $\beta\kappa=1,5,10$ and $N=642$, $t_{\max}=2\times10^8$, $R=1$, $C=2$, and $\hat \sigma= 30$.\label{snapshots}}
\end{figure}
Examples of snapshots of the vesicle for $N=642$ are shown in \fig{snapshots}. The dimensionless surface tension is $\hat\sigma=30$ and the bending modulus is varied from $\beta\kappa=1$ to 10. Note that the case $\beta\kappa=1$ corresponds more to a soap bubble than a lipidic bilayer. One observes visually that, at fixed $\hat\sigma$, the excess area decreases with increasing $\beta\kappa$ in agreement with \eq{alphaS}. Note that the roughness remains almost constant, a result which cannot be seen immediately from \eq{roughS} but becomes clear in the planar limit, \eq{roughP}.

\subsection{Height-height correlation function}

\begin{figure}[!t]
\begin{center}
\includegraphics[width=\linewidth]{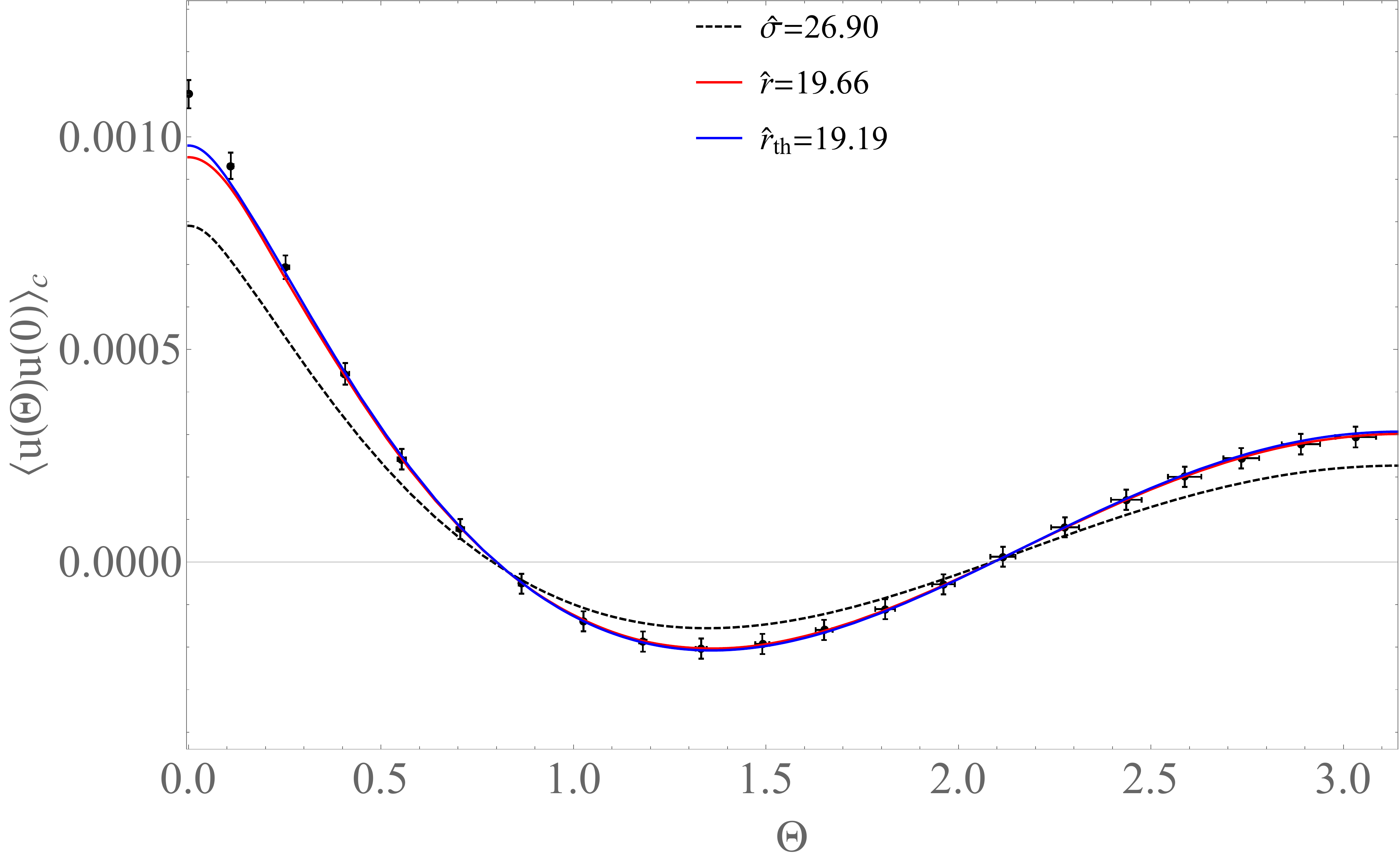}
\end{center}
\caption{Numerical angular correlation functions $\langle u(\Theta)u(0)\rangle$ for equilibrated vesicles with $\beta\kappa=10$, $\hat\sigma=26.9$, $N=642$, $R=1$, $C=0$, and $t_{\max}=5\times10^9$. Dots are the numerical results and curves corresponds to $\mathcal{C}(\Theta;\hat r)$ in \eq{CS}, for $\hat r=\hat\sigma$ (dashed line), $\hat r=19.7$ is the fitted value (red line), and $\hat \sigma_{\rm eff}=19.2$ (blue line) defined in \eq{sigma_eff_eps}. \label{correlfunctions}}
\end{figure}

The numerical correlation function $\mathcal{C}(\Theta;\hat\sigma)=\langle u(\Theta)u(0)\rangle$ is plotted in \fig{correlfunctions} for $\hat\sigma=\bar\sigma R^2/\kappa=26.9$. We observe the classical enhanced correlation between points located on opposite sites on the sphere (for $\Theta=\pi$) and an anti-correlation near $\Theta=\pi/2$ due to the excitation of the $\ell=2$ modes. Apart from the first point, the fit using \eq{CS} with $\sigma$ left as a fitting parameter, and noted $r$, the fluctuation tension, is very good. One finds $\hat r\equiv R^2 r/\kappa=19.7\pm0.4$. In~\fig{correlfunctions}, $\mathcal{C}(\Theta;\hat\sigma)$ is also shown with $\hat\sigma=26.9$ given in the input of the simulations. Hence, we observe that $r\leq\sigma$.

\subsection{Fluctuation tension}

\begin{figure}[!t]
\begin{center}
(a)\includegraphics[width=0.9\linewidth]{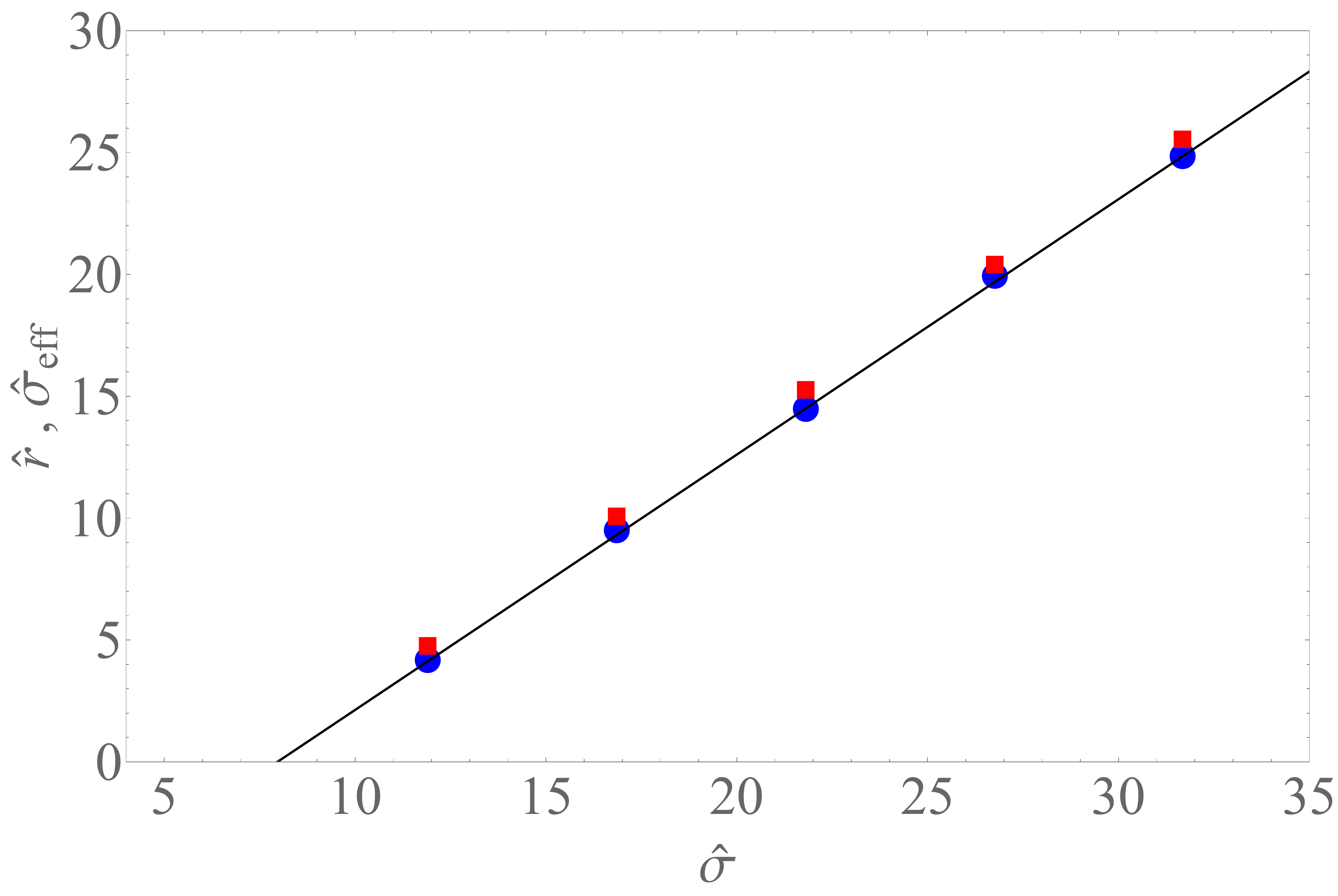}
(b)\includegraphics[width=0.9\linewidth]{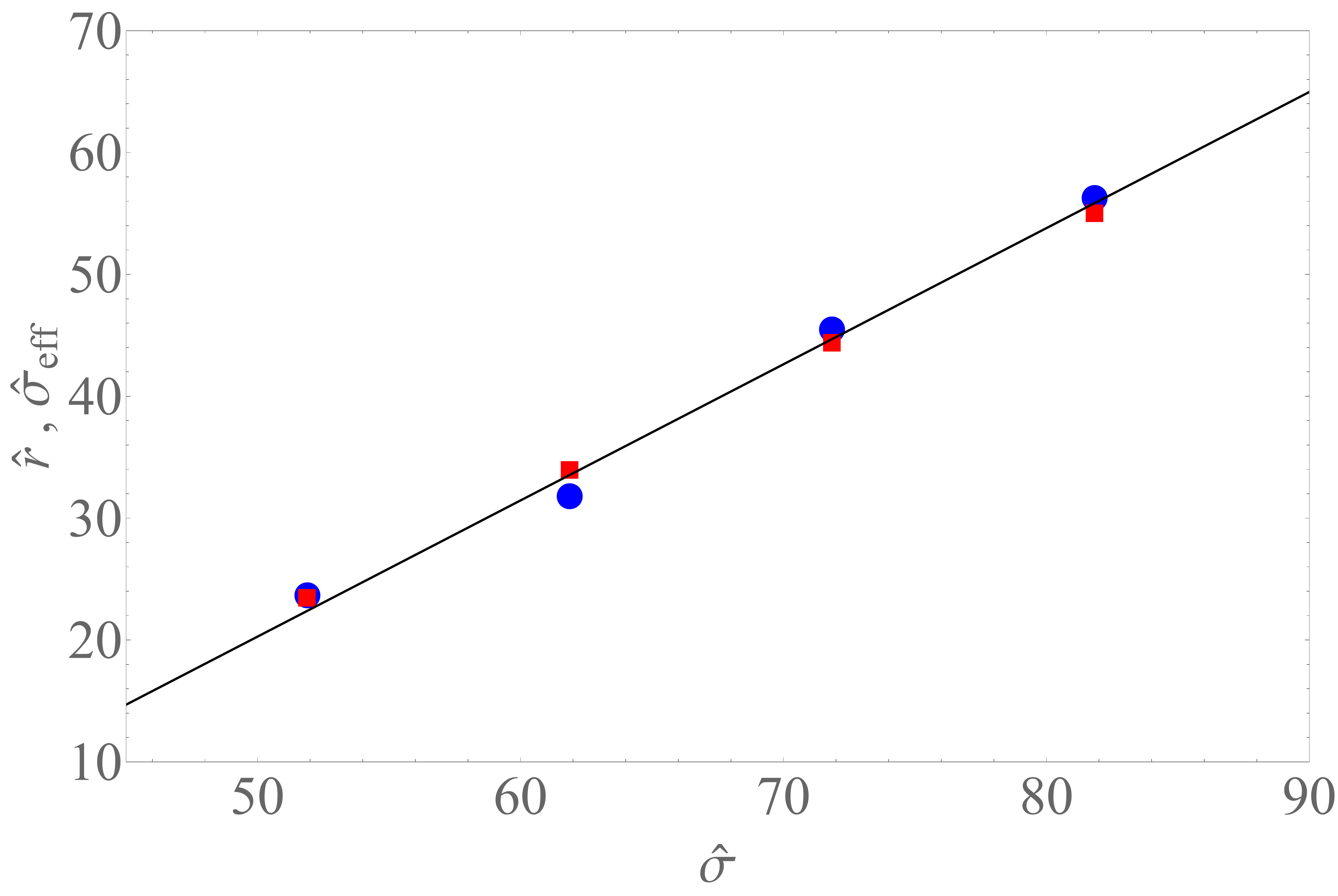}
(c)\includegraphics[width=0.9\linewidth]{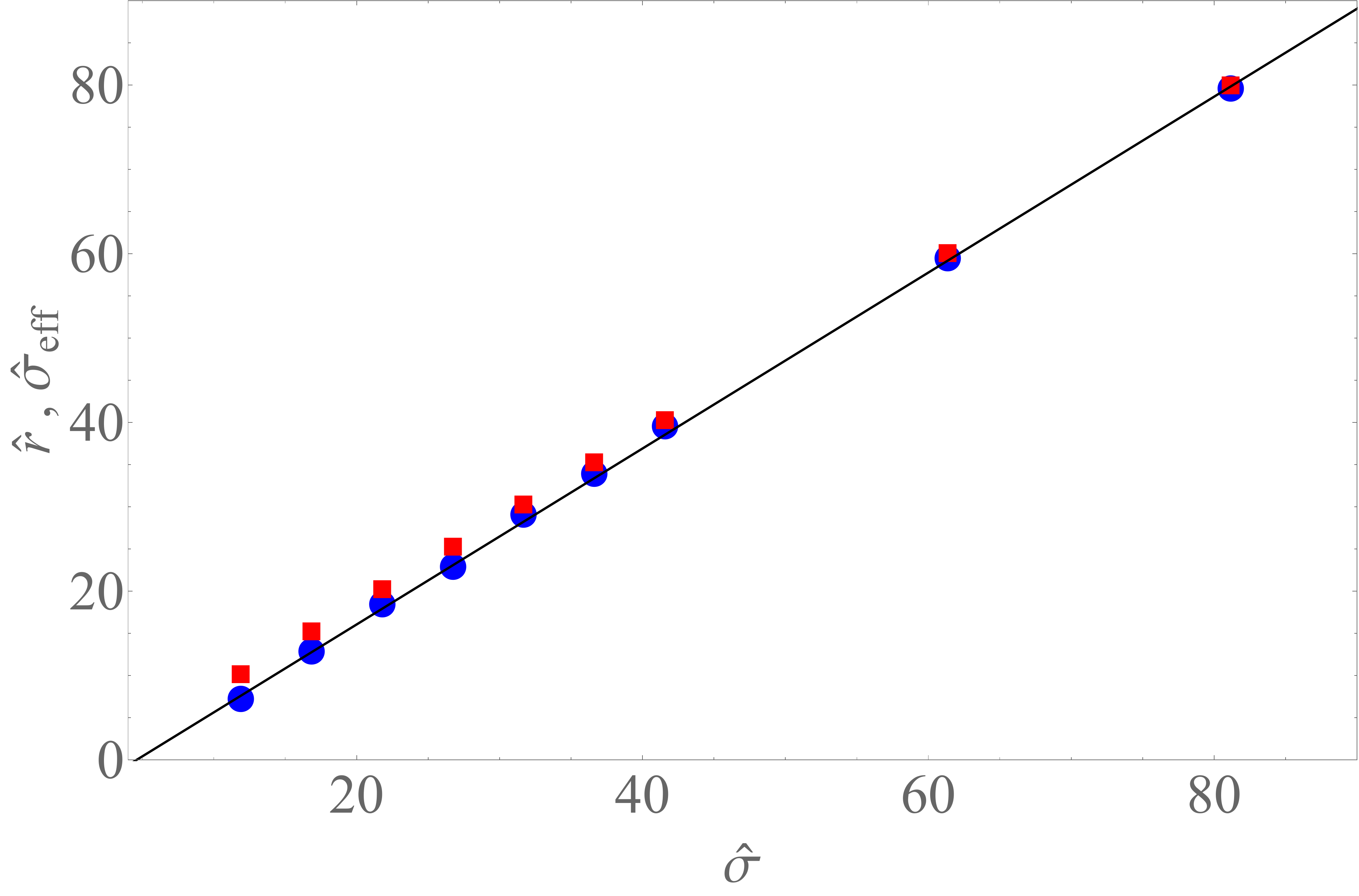}
\end{center}
\caption{Surface tension fitted from the numerical angular correlation function $\hat r$ (blue circles), and renormalized surface tension $\hat \sigma_{\rm eff}=R^2\sigma_{\rm eff}/\kappa$ given by \eq{sigmaeffNum} (red squares), vs. the bare surface tension $\hat\sigma=R^2\bar\sigma/\kappa$ for equilibrated vesicles for the same parameters as \fig{correlfunctions}. Simulations with radial MC moves (a)~$N=642$, (b)~$N=2562$, and~(c) FP corrected radial moves ($N=642$). The linear fits are (a)~ $\hat r= -8.3 + 1.0\, \hat \sigma$, (b)~ $\hat r= -35.6 + 1.1\, \hat \sigma$, and (c)~$\hat r=-4.8 + 1.0\, \hat \sigma$.\label{r=f(sigma)}}
\end{figure}
The fluctuation tension $r$ fitted in \fig{correlfunctions} is not equal to bare surface tension $\bar \sigma$. The natural other candidate is the renormalized surface tension $\sigma_{\rm eff}$ in \eq{sigma_eff_eps}. This tension has been calculated for an infinite planar membrane. In the case of finite quasi-spherical membranes studied in the simulations, three modifications must be done: i)~the system size is finite, which enforces the excess area to remain always finite. Indeed, we have measured that it remains small $\alpha\leq 5\%$ in the simulations we have performed. ii)~The bare surface tension $\sigma$ is replaced by $\bar \sigma$ since $2/R$ remains finite. iii)~The renormalisation by the factor $\Lambda^2/(8\pi\beta)=N'/(2A_{\rm p}\beta)$ comes directly from the bending fluctuations. In the quasi-spherical case we therefore decide to use the measured value of the bending fluctuation contribution $\Delta E_{\rm b}/A$, measured numerically. The computed renormalized surface tension is therefore
\be
\sigma_{\rm eff} \simeq \bar \sigma-\epsilon\frac{\Delta E_{\rm b}}{A}
\label{sigmaeffNum}
\ee
One observes for the correlation function \eq{CS}, plotted in \fig{correlfunctions} that the agreement is excellent: $\hat \sigma_{\rm eff} =19.2$ (in blue) and $\hat r=19.7$ (in red). 
\begin{figure*}[!t]
\begin{center}
\includegraphics[width=0.8\linewidth]{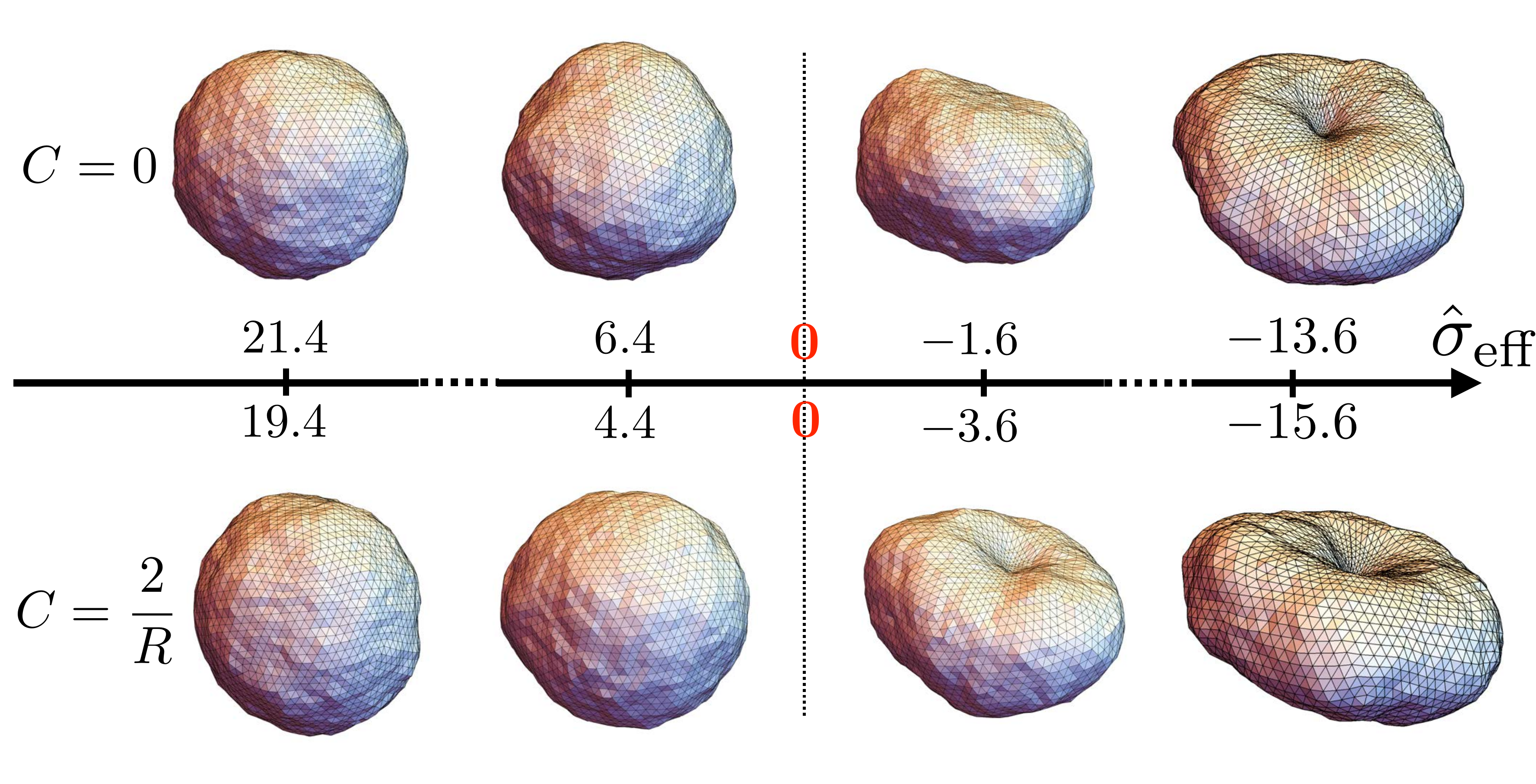}
\end{center}
\caption{Snapshots of equilibrated vesicles (after $t_{\max}=2\times10^9$~MC steps) for $C=0$ (top) and $C=2/R$ (bottom), for positive and negative values of $\hat\sigma_{\rm eff}$, corresponding to positive values of $\hat\sigma$ ($N=2562, \beta\kappa=10, R=1$). \label{shapes}}
\end{figure*}

In \fig{r=f(sigma)}, the comparison between the dimensionless fitted fluctuation tension $\hat r\equiv R^2 r/\kappa$ and $\hat \sigma_{\rm eff}\equiv R^2 \sigma_{\rm eff}/\kappa$ is done for various values of $\sigma$ and two values of $N$, $N=642$ and $N=2562$ by repeating the same fitting procedure.

These tensions, $\hat r$  (solid blue circles) and $\hat \sigma_{\rm eff}$  (solid red squares) are plotted against $\hat \sigma\equiv R^2 \bar\sigma/\kappa=R^2 \sigma/\kappa+ 2$ (since $C=0$), and for $\beta\kappa=10$ in \fig{r=f(sigma)}.  Figures~\ref{r=f(sigma)}(a) and~(b) correspond to radial MC moves with $\epsilon=3$. 
To check \eq{sigmaeffNum} with the FP correction, we also perform simulations with corrected radial moves following \eq{FP_dx}, that is by keeping constant the length of the MC move along the normal ($\epsilon=1$). The resulting fluctuation tension is plotted in \fig{r=f(sigma)}c for $N=642$. 

We clearly see that the fluctuation tension $r$ matches with the renormalized surface tension $\sigma_{\rm eff}$ with a very good agreement, and both vary linearly with $\sigma$ as expected from \eq{sigma_eff_eps}. This is true both with or without the correction taking into account the Faddeev-Popov term. Indeed the extrapolated intercept is 8.3 and 32.4 respectively for $N=642$ and 2562 with radial moves to be compared to the analytical value of the residual tension $\hat\sigma_c= 3N/(8\pi\beta\kappa)$ equal to 7.7 and 30.6. When corrected radial FP moves are included in the simulation, the comparison is worse, 4.8 vs. 2.6. Indeed close to the transition, due to large fluctuations the correction given in~\eq{FP_dx} is less adapted and real moves along $\bn$ would be more accurate.
 
\subsection{Vesicle shape transition}

For infinite planar membranes, the excess area $\alpha_{\rm pl}$ diverges following \eq{alpha_eff_cutoff} at the transition, $\sigma_{\rm eff}=0$. But what happens for vesicles of finite size? 
Seifert~\cite{seifert1995} has shown that for quasi-spherical vesicles and in the limit $\beta\kappa\alpha\gg 1$ (\textit{i.e} for low $\sigma$), the modes for $l=2$ pick up most of the excess area which is therefore the signature of a change of shape.

From \fig{r=f(sigma)}(b) ($N=2562$, $\beta\kappa=10$, radial MC moves $\epsilon=3$), one can predict the critical bare surface tension $\hat\sigma_{c}$ such that $r$ (or $\sigma_{\rm eff}$) vanishes. Using the linear fit, one finds $\hat\sigma_{c}=32$ indeed very close to the expected value from \eq{sigma_eff_eps}, $\sigma_c\simeq 3N/(8\pi\beta\kappa)= 31$.
Note that since $\epsilon=3$ in this case, the Laplace and frame tensions do not vanish at $\sigma_{c}$ but $\gamma\simeq \tau\simeq k_{\rm B}TN/A_{\rm s}$ (since $A\simeq A_{\rm s}$ at zeroth order in $\alpha\leq 5\%$). 

We have measured (data not shown) that the correlation time decreases with the fluctuation surface tension like $\tau_{\rm corr}\propto r^{-\lambda}$ with $\lambda$ an apparent exponent which depends on $N$ and is close to 0.6 for $N=642$ and 1.1 for $N=2562$. Therefore it is extremely long to simulate a vesicle exactly at the transition and we show in Fig.~\ref{shapes} snapshots of equilibrated vesicles on simulation times  $t_{\max}\simeq 10\ \tau_{\rm corr}$ just below and above this transition.

By inspecting the snapshots for decreasing values of $\hat \sigma$ defined in \eq{sigmahat} both for $C=0$ and $C=2/R$, one clearly observes that the mean vesicle shape remains spherical for $\hat\sigma_{\rm eff}>0$ ($\hat\sigma>\hat\sigma_{c}$), which corresponds to the tension regime.
For values of $\sigma_{\rm eff}\gtrsim0$ the mean shape remains spherical but with large fluctuations, for both values of $C$, thus confirming that the spontaneous curvature solely enters in the surface tension. 
However, for lower $\hat\sigma_{\rm eff}<0$ ($0<\hat\sigma<\hat\sigma_{c}$), the equilibrated vesicle shape is no more spherical. There is a shape transition where a symmetry is broken along at least one direction.

Once the effective membrane tension becomes negative, its effect is now to increase the excess area. The vesicle shape is controlled by the balance between the bending energy, \eq{H0}, and the negative surface tension. The expansion around a sphere leading to the results of Section~\ref{Gauss} is no more valid after the shape transition. Indeed we observe in \fig{shapes} two typical different shapes according to the values of $C$. For $C=0$ (top) two quasi-planar surfaces are formed whereas for $C=2/R$ (bottom) the vesicle has essentially a positively curved  region which satisfies $C\simeq 2/R$ at the expense of a smaller region with a negative curvature. This negatively curved region allows the vesicle to store the excess area. 
The vesicle wants to maximise its surface area while keeping the volume constraint. Moreover it still wants to minimise its bending energy since $\beta\kappa\simeq 10-50$ ($\kappa\simeq4-20\times 10^{-20}$~J~\cite{pecreaux} for usual lipid bilayers). Hence the wave-vector which destabilises the membrane is $q_c\simeq\sqrt{-\sigma_{\rm eff}/\kappa_{\rm eff}}$ (for infinitely large membranes, the membrane is mechanically unstable). One typically obtains $q_c^{-1}\simeq R/2$ which is in agreement with the snapshots shown in~\fig{r=f(sigma)}.
If we decrease $\sigma_{\rm eff}$ further, we obtain vesicles having the shape of a ``donut'' (with negatively curved regions in the center on both sides but without any hole) whatever the value of $C$ (see \fig{shapes}), thus confirming that the shape does not depend on $C$ anymore. For even lower values of $\sigma_{\rm eff}$ (but still with $\sigma>0$), several lobes eventually appear to increase further the area (not shown).

We are now able to propose a new interpretation of the experimental results on shape transitions obtained by K\"as and Sackmann in 1991~\cite{kas}. Vesicles of typical radius of 15~$\mu$m are observed at various temperatures. In particular in their Fig.~3, the vesicle changes from a quasi-spherical shape at $T=27.2~^\circ$C to a prolate ellipsoidal one at $T=36~^\circ$C with large planar domains similar to the vesicle shape shown in \fig{shapes} for $C=0$ and $\hat\sigma_{\rm eff}=-1.6$. For higher temperatures a budding transition is observed. Using our \eq{sigma_eff_eps}, one can roughly estimate the effective surface tension at $T_1=27.2~^\circ$C assuming that the transition $\sigma_{\rm eff}=0$ occurs for $T_2=36~^\circ$C (we suppose $\sigma$ independent of $T$):
\be
\sigma_{\rm eff}(T_1)=\sigma_{\rm eff}(T_2)+\frac{k_{\rm B}(T_2-T_1)}{2a_p}
\ee
which leads to $\sigma_{\rm eff}(T_1)=2.4\times 10^{-6}$~N/m for $a_p=25~\mathrm{nm}^2$, a reasonable value compared to surface tensions measured in experiments~\cite{pecreaux,schmidt,monzel}.

\section{Discussion and conclusions}
\subsection{Discussion of previous simulations of planar membranes}

In light of the preceding results on simulated vesicles, we are able to discuss the numerical results presented by Shiba \textit{et al.} in a recent paper on planar membranes~\cite{shiba}.
In this article, the authors measure the fluctuation surface tension $r$ as a function of the applied frame tension $\tau$ using corrected MC moves (equivalent to $\epsilon=1$). Note that they work in the $(\tau,A)$ ensemble, \textit{i.e.} the projected area $A_{\rm s}$ is allowed to fluctuate, and the true area $A$ is kept constant by adjusting $\sigma$. Since they have fixed $N=6400$, modifying $\tau$ is equivalent to modifying $a_{\rm s}=A_{\rm s}/N=4\pi/\Lambda^2$. They adjust $\sigma$ and measure the fluctuation tension $r$ by fitting the fluctuation spectrum. Since $N$ is large, the finite size effects are negligible and the ensemble $(\tau,A)$ that they study can be considered as equivalent to the ensemble $(A_{\rm s},\sigma)$. We have shown in Section~\ref{Gauss}, that the ensemble $(A_{\rm s},\sigma)$ corresponds to the $(V,\sigma)$ ensemble for vesicles that we considered in this work after taking the thermodynamic limit $N\to\infty$ and $R\to\infty$ while keeping $a_{\rm s}$ constant. According to \eqs{taur}{gamr}, the fluctuation tension $r\simeq \tau$ since they have measured $\alpha\leq3\%$ in their simulations and we can therefore assume $\gamma\simeq\tau$.
We assume that in the $(\tau,\sigma)$ ensemble, $\tau$ is not (or slightly) renormalized, and corresponds to $\tau_{\rm eff}$ in our ensemble. 

Interestingly they find that when the applied frame tension vanishes, $\tau=0$, the bare surface tension $\sigma=\frac12k_{\rm B}T/a^2=N/(2\beta A)$ which is almost (by a factor $1+\alpha\simeq1.03$) our residual tension, $\sigma_c=k_{\rm B}T\Lambda^2/8\pi$. This is in full agreement with our \eq{taur} which for $\epsilon=1$ writes $\tau_{\rm eff}=\sigma_{\rm eff}(1+\alpha_{\rm eff})$ hence $\tau_{\rm eff}(\sigma_c)=0$.

Moreover, Shiba \textit{et al.}~\cite{shiba} measure three different fluctuation tensions: $r$ which corresponds to the full Hamiltonian with the corrective term \eq{shiba_corr}, and $r_{\rm EL}$ (respectively $r_{\rm L}$) which corresponds to a truncated Helfrich Hamiltonian at order 2 in $h$ with (resp. without) the corrective term of \eq{shiba_corr}. These three fluctuation tensions are plotted in their fig.~5.  First they obtained $(\sigma-\tau)/\sigma\simeq 1$ for $\tau=10^{-2}$ (in units of $k_{\rm B}TN/A$). Following \eq{sigmaeffNum} this implies that $\Delta E_{\rm b}/A_{\rm s}\simeq \sigma$. Furthermore they obtained $r=\tau$ for all the range of $\tau$ values, which confirms our result plotted in \fig{r=f(sigma)}c, that $r=\sigma_{\rm eff}$. Finally, they measured $r_{\rm EL}-\tau\simeq 3 (\sigma-\tau)$. Indeed, the Gaussian fluctuation tension is $\sigma$ and the FP correction brings the additional term $2\Delta E_{\rm b}/A_{\rm s}$. Thus one has $r_{\rm EL}-\tau=\sigma +2\Delta E_{\rm b}/A_{\rm s}-\tau\simeq 3\Delta E_{\rm b}/A_{\rm s}$. Using their result $\Delta E_{\rm b}/A_{\rm s}\simeq \sigma$, one finds $(r_{\rm EL}-\tau)/\sigma\simeq 3$ as observed.

Finally, they observe that for $\tau=0$ the membrane remains flat on average. This is also in agreement with our results. Indeed for a planar membrane, $C=0$. At the transition, such that $-1\lesssim\hat\sigma_{\rm eff} <0$, the shape is only controlled by the bending energy which favours a flat membrane on average. It corresponds to the top-left snapshot of \fig{shapes} in the vesicle case. This is the reason why they do not observe any shape transition in their simulations for $\tau$ close to 0.

Similar simulations of planar membranes have been done previously by Avital and Farago~\cite{avital} at fixed frame tension $\tau$.
By applying a weak $\tau$, they measured $r$ and the two areas $A$ and $A_{\rm p}$. Contrary to~\cite{shiba}, they do not control $\sigma$ which comes from the force field that they use for the interaction between lipids in their MC simulations. They observed that $\tau\simeq r$ for large surface tensions. This is in agreement with our \eq{taur} if we assume that they use normal moves in their MC simulations (which is not specified).
Imparato~\cite{imparato} simulated a planar membrane using Molecular Dynamics, by controlling $A_{\rm p}$. He then measured $r$, $A$ and $\tau$ (by computing the stress tensor). He also observed that the law $r\simeq\tau$ is verified for $\tau>0$.
 
In these three numerical studies, the fluctuation tension $r$ was also measured for negative $\tau$. Avital and Farago observe that the fluctuation tension $r$ saturates at a small but negative value, whereas Imparato finds $r\simeq 0$ for $\tau<0$. Shiba \textit{et al.} did not measure $r$ but observed that the planar membrane buckles when they decrease further $\tau$ to a critical negative value $\tau_b<0$. This is equivalent, in the $(A_{\rm p},\sigma)$ ensemble, to create some area $A$ (indeed they have a ``negative" surface tension $\sigma$) by maintaining $A_{\rm p}$ fixed.  Hence, as in our vesicle simulations shown in \fig{r=f(sigma)} for negative $\sigma_{\rm eff}$, the membrane buckles to increase its area but does not crumple because $\beta\kappa=10$ remains large.

Note that in our simulations, we did not compute the fluctuation tension $r$ of vesicles for values of $\sigma<\sigma_c$ (such that $\sigma_{\rm eff}<0$) since the theoretical correlation function of a fluctuating vesicle with a non quasi-spherical shape is not known and the correlation time is very large close to the transition. 

We emphasise that the equality (or difference) between the Laplace or frame tension and the fluctuation tension cannot be shown analytically within the Gaussian approximation as tried in several papers~\cite{fournierPRL,barbetta,schmid,deserno}. Moreover, extracting some physical information when the Laplace tension (as in~\fig{gammag}) or frame tension (as in~\cite{barbetta}) vanish (or become negative) using the Gaussian approximation may lead to some erroneous conclusions. 

In recent MC simulations on fluctuating vesicles, using a tethered network model similar to the one developed in Ref.~\cite{gompper96} where local constraints are applied on the bond length $\zeta$ but no global constraint on the area is added, Peni\v{c} \textit{et al.}~\cite{penic} have measured a slightly negative effective surface tension $-2.4\leq \hat\sigma_{\rm eff}\leq0.2$. According to our results, this would mean that their vesicles are very close to the shape transition. Hence the surface tension induced by the local constraints is close to the residual tension $\sigma_c=\epsilon k_{\rm B}T\pi/\zeta$. Further quantitative analyses on such model would be interesting to confirm this result.

\subsection{Summary}

In this paper, we have studied the different surface tensions of closed vesicle membranes, namely the Laplace tension $\gamma$ appearing in the Laplace equation \eq{Laplace}, and the fluctuation tension $r$. To do so we have developed the statistical physics of the membrane using the Helfrich Hamiltonian at the Gaussian level, valid for highly taut membranes, and using the renormalisation group calculations, necessary when the surface tension decreases and the membrane roughness becomes important. We then have successfully compared our analytical predictions to numerical results obtained using extensive Monte Carlo simulations ($N$ up to 2500, $t_{\rm sim}\simeq 10^3\ \tau_{\rm corr}$) of tessellated vesicles.

In the analytical part, we have computed $\gamma$ and the ``mechanical'' frame tension $\tau=\gamma/(1+\alpha)$, initially introduced for planar membranes supported on a frame, as a function of the bare surface tension $\sigma$, the bending modulus $\kappa$ and the spontaneous curvature $C$. We have shown that for large vesicles, these expressions are similar to the ones obtained for planar membranes which led us to study the renormalisation of $\sigma$, $\tau$ and $\gamma$ in the planar case. 
We have properly shown that the bare surface tension is renormalised following \eq{sigma_eff_eps}, where $\epsilon=1$ or 3 according to the allowed membrane moves (along the local normal $\bn$ to the membrane or along the averaged one, $\mathbf{e}_z$) in the renormalisation procedure. This is a different result as compared to earlier works~\cite{peliti85,meunier87,david91}. This effective surface tension $\sigma_{\rm eff}$ therefore corresponds to the fluctuation surface tension measured experimentally by fitting the fluctuation spectrum. We then have deduced the renormalised Laplace and frame tensions, and shown that in the case where $\epsilon=1$, the former is equal to the renormalised surface tension $\sigma_{\rm eff}$.

In the numerical work, we developed a numerical vesicle model without the introduction of any local harmonic spring to enforce the patch sizes, as usually done in previous numerical studies. We applied more physical global constraints: a fixed volume and either a fixed total area or a fixed bare surface tension $\sigma$. This is the most reliable way to model numerically the Helfrich Hamiltonian and therefore to do a direct comparison.
For quasi-spherical vesicles with a small excess area $\alpha$, we have shown an excellent agreement between the numerical roughness, excess area and averaged curvature energy and the analytical expressions. We have compared the fluctuation surface tension $r$, fitted from the correlation function, to $\sigma_{\rm eff}$.
We have shown that the simple formula, \eq{sigma_eff_eps}, computed using the renormalisation procedure of the surface tension of planar membranes, is in excellent agreement with the fluctuation tension for the simulated vesicles for a broad range of values of $\sigma$ and for the two types of MC moves: radial moves or corrected radial moves to mimic normal ones. To do so we have measured the average of the bending energy, noted $\Delta E_{\rm b}/A_{\rm s}\simeq k_{\rm B}TN/A_{\rm s}$. For radial moves, one finds $r=\sigma -3k_{\rm B}TN/A_{\rm s}$ whereas for corrected radial (``normal'') moves $r=\sigma -k_{\rm B}TN/A_{\rm s}$. 

Furthermore, for bare surface tensions close to the residual tension $\sigma_c$ such that $\sigma_{\rm eff}<0$, one observes a shape transition, the vesicle changing from a mean spherical shape to an oblate or prolate shape. The vesicle then shows flat or negatively curved domains, depending on the value of the spontaneous curvature $C$. We argued that this transition is the signature in the vesicle case of the famous flat to crumpled thermodynamic phase transition for infinite flat membranes~\cite{membrane_book}, corresponding to the vanishing of $\sigma_{\rm eff}$ and the divergence of $\alpha_{\rm pl}$. But for finite systems such as the ones explored in the simulations with $\kappa\simeq 10k_{\rm B}T$, the renormalised bending modulus remains larger than $k_{\rm B}T$ and the membrane physics is controlled by the balance between the bending energy and the negative surface tension, which tends to increase its area, without diverging however. Hence the various deformed shapes can be seen as buckled shapes. The rich diversity of vesicle shapes then comes from all the possible values of the spontaneous curvature. Moreover our result \eq{sigma_eff_eps} explains why one can have $\sigma_{\rm eff}<0$ but still keeping $\sigma>0$, which is mathematically necessary to have definite Gaussian integrals in the Helfrich theory.

\subsection{Outlooks}

The study of the shape transition is not fully quantitative for several reasons.
Close to the transition $\sigma_{\rm eff}\simeq 0$, the correlation time becomes very large and we are faced with a classical slowing down issue in statistical physics. One way to circumvent this would be to do a thorough finite-size study of the simulations. Moreover, we use corrected radial MC moves, \eq{FP_dx}, and another way to improve the localisation of the transition would be to modify the numerical code by introducing exact normal MC moves exactly along $\bn$. In the future, we plan to do a more systematic exploration of the physical parameter space and in particular to elucidate further the role played by the spontaneous curvature $C$. Although for quasi-spherical vesicles it only modifies the surface tension, at the transition it controls the vesicle shape, as shown in \fig{shapes}.

To focus on the Laplace surface tension, it would be interesting to work in the $(\sigma, \Delta P)$ ensemble, where the vesicle volume is controlled by a pressure difference $\Delta P$. This ensemble would correspond to the $(\sigma,\tau)$ ensemble for planar membranes.

Importantly, understanding theoretically the fluctuation tension and its renormalisation is a pre-requisite to perform detailed theoretical~\cite{leibler1,schick,dean,GueguenEPJE} and numerical~\cite{bagatolli,amazon2013,amazon2014} studies of multicomponent membranes. Indeed, one should expect that any lipid phase separation in the membrane controlled by the lipid curvature would play a role on the membrane shape through a modification of the surface tension.

\acknowledgments
We thank Julie Cornet for her valuable simulation work as part of her Master's project.

\end{document}